\documentclass[a4paper,11pt]{article}
\usepackage{jcappub}
\usepackage{lineno}
\usepackage{graphicx}
\usepackage[utf8,latin1]{inputenc}
\usepackage{dcolumn}
\usepackage{mathrsfs}
\usepackage{subcaption}
\usepackage{caption}
\usepackage{academicons}
\usepackage{mathtools, nccmath}
\usepackage{tensor}
\usepackage{comment}
\usepackage[overload]{empheq}
\usepackage{tikz,xcolor}
\usepackage{cases}
\usepackage{setspace}
\usepackage{lipsum}
\usepackage[normalem]{ulem}
\usepackage{bm}
\usepackage{soul}
\usepackage{hhline}
\usepackage{bigints}
\usepackage{hyperref}
\usepackage{amsmath}
\usepackage[T1]{fontenc}
\usepackage{fancyhdr}
\newcommand{\qm}[1]{``#1''}
\usepackage{hyperref}
\hypersetup{colorlinks=true, linkcolor={blue},citecolor={blue},urlcolor={blue}}  

\newcommand{\bsubeqs}{\begin{subequations}}
\newcommand{\esubeqs}{\end{subequations}}

\definecolor{lime}{HTML}{A6CE39}
\DeclareRobustCommand{\orcidicon}{
	\begin{tikzpicture}
	\draw[lime, fill=lime] (0,0) 
	circle [radius=0.16] 
	node[white] {{\fontfamily{qag}\selectfont \tiny ID}};
	\draw[white, fill=white] (-0.0625,0.095) 
	circle [radius=0.007];
	\end{tikzpicture}
	\hspace{-2mm}
}
\newcommand{\dd}{{\rm d}}
\newcommand{\ee}{{\rm e}}
\newcommand{\OO}{{\rm O}}

\newcommand{\RS}{R_{\rm S}}

\fancypagestyle{plain}{%
  \fancyhf{}
  \fancyfoot[C]{\iffloatpage{}{\thepage}}
  }
\foreach \x in {A, ..., Z}{%
	\expandafter\xdef\csname orcid\x\endcsname{\noexpand\href{https://orcid.org/\csname orcidauthor\x\endcsname}{\noexpand\orcidicon}}
}

\begin{document}

\title{Energy conditions in static, spherically symmetric spacetimes and effective geometries}

\author[a]{Zi-Liang Wang\orcidA{}}
\author[b]{Emmanuele Battista\orcidB{}}

\affiliation[a]{Department of Physics, School of Science, Jiangsu University of Science and Technology, Zhenjiang, 212003, China}
\affiliation[b]{Istituto Nazionale di Fisica Nucleare, Laboratori Nazionali di Frascati, 00044 Frascati, Italy}

\emailAdd{ziliang.wang@just.edu.cn}
\emailAdd{ebattista@lnf.infn.it,emmanuelebattista@gmail.com}

\abstract{Classical energy conditions are investigated in generic static and spherically symmetric spacetimes. In setups with  nonconstant $g_{tt}  g_{rr}$, the appearance of horizons can signal the violation of the  null energy condition  and the breakdown of some standard near-horizon properties. For configurations satisfying $g_{tt}g_{rr}=-1$, we devise a systematic algorithm to generate solutions of the Einstein field equations that automatically obey the null energy condition. Within this family, we select a particularly significant metric that incorporates a logarithmic correction to the Schwarzschild model and fulfills all standard energy criteria. We examine its main features, including the horizon structure, geodesic behavior, and junction conditions. Our analysis shows that this geometry can be interpreted as an effective exterior description for both horizon-bearing and  horizonless compact objects, and suggests that it can  potentially act, in certain regimes, as a black hole mimicker.  }

\arxivnumber{2604.16545}
\maketitle
\flushbottom

\section{Introduction}
\label{sec:intro}


Einstein equations relate spacetime curvature to the matter content, but if no suitable restrictions are imposed on the energy-momentum tensor  $T_{\mu \nu}$, then, in principle, solutions can be generated \emph{ad libitum} and geometries lacking any physical foundation are likely to be encountered. This undesirable situation can be avoided by enforcing  classical energy conditions  \cite{Hawking1973-book,Geroch1979,visser1995lorentzian,Poisson2009,Martin-Moruno:2017,Curiel2017,Kontou2020,Martin-Moruno:2013wfa}, which consist of pointwise inequalities requiring that various linear combinations of the components of  $T_{\mu \nu}$ take non-negative values.  In this way, the notion that   \qm{normal} matter behaves in a physically reasonable  way, i.e., consistently with classical expectations and such that  gravity remains attractive, is implemented.

The weakest constraint on $T_{\mu \nu}$ is the so-called  null energy condition (NEC), which asserts that, for any null four-vector $k^{\mu}$, the following inequality must hold:
\begin{align}
T_{\mu \nu} k^{\mu} k^{\nu} \geq 0 \,,\label{NEC-vectors}
\end{align}
while the weak energy condition (WEC) is  expressed as
\begin{align}\label{eq:def_v}
T_{\mu \nu} v^{\mu} v^{\nu} \geq 0 \,,
\end{align}
with $v^{\mu}$ any future-directed, normalized, timelike four-vector. The WEC implies the NEC by continuity and  demands that the energy density locally measured by an  observer with four-velocity $v^{\mu}$ be non-negative, a property typically regarded as characteristic of \qm{ordinary} (i.e., \qm{non-exotic}) classical matter \cite{Barcelo2000}.

The assumption that gravity be always attractive in general relativity can be formulated via the strong energy condition (SEC) \cite{Martin-Moruno:2017}, which entails that
\begin{align}
\left(T_{\mu \nu} - \frac{1}{2}T\, g_{\mu \nu}\right) v^{\mu} v^{\nu} \geq 0 \,, 
\end{align}
and clearly guarantees the NEC, but not necessarily  the WEC.

Finally, the dominant energy condition (DEC) stipulates that no experimenter detects a negative energy density and that the associated energy-momentum flux $J^\mu=-T^{\mu}{}_{\nu} v^{\nu}$ be future-directed and non-spacelike, ensuring that  energy never propagates superluminally. These prescriptions amount to
\begin{align}
T_{\mu \nu} v^{\mu} v^{\nu} \geq 0, \qquad J^\mu J_\mu \leq 0 \,,
\end{align}
and one can immediately see  that DEC yields the WEC and hence the NEC.

Several foundational classical results rely on the validity of the above  energy criteria, including  the singularity \cite{Penrose:1964wq,Hawking1967,Hawking:1970zqf},   positive mass \cite{Schoen1979a,Schoen1979b,Horowitz1984}, and  censorship  \cite{Olum1998,Visser1998a} theorems, as well as  black hole thermodynamics \cite{Wald,Wald-LRR} and certain types of no-hair theorems \cite{Mayo1996,Herdeiro2015}  (see e.g. Ref. \cite{Curiel2017} for further details). In particular, the 1965 Penrose theorem establishes the null geodesic incompleteness of spacetimes such as Schwarzschild, and rests on  the null convergence condition $R_{\mu \nu} k^\mu k^\nu \geq 0$, which is equivalent to the NEC in the context of general relativity (see e.g. Refs. \cite{Borissova2025b,Borissova2025}). In addition, the 1970 Hawking-Penrose theorem employs the stronger timelike convergence condition $R_{\mu \nu} v^\mu v^\nu \geq 0$, tantamount to the SEC within Einstein gravity, and holds for solutions such as Reissner-Nordstr\"{o}m, where the original Penrose result is inapplicable due to the presence of a Cauchy horizon that breaks global hyperbolicity. The advantage of convergence conditions is that they configure as general geometric assumptions, which do not entail a particular choice of dynamical equations.

On the other hand, energy conditions are not fulfilled by many quantum fields (most famously in the Casimir effect and Hawking evaporation \cite{visser1995lorentzian}), and even by  some simple classical systems  (in particular, the non-minimally coupled scalar fields \cite{Barcelo2000}, while minimally coupled massive scalar fields  fail to obey the SEC \cite{visser1995lorentzian}); a detailed list of  classical and quantum violations can be found in Refs. \cite{Curiel2017,Kontou2020}. In response to these challenges, weaker statements are introduced: quantum energy inequalities  provide (state-independent) lower bounds on weighted averages of $\langle T_{\mu \nu} \rangle_{\rm ren}$,  while the averaged energy conditions  involve integrals  of $T_{\mu \nu}$ along suitable causal curves. The latter are  weaker than the ordinary pointwise constraints and thus take an intermediate position between the classical criteria   and the quantum  ones \cite{Ford1994,visser1995lorentzian,Fewster2012,Martin-Moruno:2017,Kontou2020}. 

One of the most significant settings restricted by standard energy conditions concerns the so-called exotic spacetimes \cite{Kontou2020}, with traversable wormholes providing a prominent example \cite{visser1995lorentzian,Lobo2017}.  Wormholes belong to the broad class of exotic compact objects and provide an example of black hole mimickers \cite{Lemos2008,Carballo-Rubio2018,Cardoso-Pani2019,Bouhmadi-Lopez:2021zwt,Biswas:2022wah,Yang2023,Biswas:2023ofz,Casadio2024,Bambi2025,DeLaurentis2025,Carballo-Rubio2025-review,Guo2026}. Such configurations are theoretically predicted in various extended theories of gravity, but also arise in certain  general-relativistic scenarios (such as patterns involving beyond-the-standard-model fields minimally coupled to gravity, or exotic states of matter), and include  fuzzballs \cite{Mathur2005,Skenderis2008}, gravastars \cite{Rahaman2014,Mottola2023,Jampolski2025a}, boson stars \cite{Liebling2012,Shnir2022}, scalarized black holes \cite{Chew:2024evh}, and several other proposals (see Table 1 in Ref. \cite{Cardoso-Pani2019} for a complete taxonomy of exotic compact objects).

Reflecting the \emph{classical} viewpoint that spacetimes failing to comply with energy conditions are often regarded as problematic, numerous  models  have been developed in which the necessary exotic matter can be minimized in suitable regimes. This occurs, for example, in some wormhole and   gravastar configurations  (see e.g. Refs. \cite{Visser1989,Visser2003a,Kar2004,DiGrezia2017,Mehdizadeh2024,Battista2024-WH,Horvat2007,MartinMoruno2012}), as well as for regularized black hole constructions known as \qm{black-bounce} spacetimes \cite{Simpson2018a,Lobo2020},  in which the singularity resolution is associated with the defocusing of null geodesics on a wormhole-throat-like surface that violates the aforementioned null convergence condition (see e.g. Ref. \cite{Borissova2025b} for details).

Motivated by this situation, in this paper we study classical energy conditions in static and spherically symmetric geometries within  general relativity. We show that, in scenarios with nonconstant $g_{tt}g_{rr}$, horizons characterized by  $g_{tt} g_{rr} =0$ can be  associated with NEC violations.  Moreover, in frameworks with $-g_{tt}=g_{rr}^{-1}$,   we propose a systematic algorithm to construct metrics  that adhere to the NEC. This procedure allows us to single out a particularly interesting case involving a logarithmic modification of the Schwarzschild  model. This solution satisfies all energy conditions and can constitute  an effective  \emph{geometric} candidate for  describing  the exterior spacetime of compact bodies, whether or not they carry an event horizon.

The structure of the paper is as follows. Classical energy criteria are  analyzed  in Sec. \ref{Sec:Static-spheric-symm-geo}. The logarithmically corrected spacetime is thoroughly examined in Sec.  \ref{Sec:geometry-log-correction}, and concluding remarks are given in Sec. \ref{Sec:conclusions}. Supplementary material is provided in the appendix. 

\emph{Conventions.} We use   $G = c = 1$ units, and  metric signature $(-+++)$.

\section{Static and spherically symmetric geometries}\label{Sec:Static-spheric-symm-geo}

In this section, we perform a detailed examination of classical energy conditions in the most general  static and  spherically symmetric spacetime 
\begin{align}\label{eq:metric_static}
{\rm d}s^2=g_{\mu \nu}{\rm d}x^\mu{\rm d}x^\nu= - B(r) {\rm d}t^2  + A(r) {\rm d}r^2 + r^2 {\rm d}\Omega^2\,,
\end{align}
with ${\rm d}\Omega^2 = {\rm d}\theta^2 + \sin^2 \theta \; {\rm d}\phi^2$\, the  line element on the two-sphere.

Energy criteria are most conveniently discussed in the local proper reference frame ($\hat{t}$, $\hat{r}$, $\hat{\theta}$, $\hat{\phi}$) of a static observer~\cite{Wald,Poisson2009}. For this reason, we introduce the non-coordinate orthonormal basis in the standard manner \cite{MTW,Nakahara2003}:
\begin{subequations}
\label{eq:tetrads}
\begin{align}
e_{a}=e_a{} ^{\mu}\, (\partial /\partial x^{\mu}),   
\end{align}
\text{with the  dual basis fulfilling}
\begin{align}
 e^a=e^a{}_{\mu} \dd x^{\mu}\,,   
\end{align}
\end{subequations}
where Latin indices $a,b,\dots=\hat{0},\hat{1},\hat{2},\hat{3}$ denote frame components,  Greek indices correspond to coordinate components, and  $e_a{}^{\mu}$ represents the inverse of $e^{a}{}_{\mu}$. These objects,   referred to as tetrads (or vielbeins),   form a $4\times 4$ matrix with  positive determinant and satisfy the orthonormality condition
 \begin{align}\label{eq:orthonormality condition}
 g_{\mu \nu}\,e_{a}{}^{\mu}e_{b}{}^{\nu}= \eta_{ab} \,,
\end{align}
as well as the completeness relation
 \begin{align}
  g_{\mu \nu} =e^{a}{}_{\mu}e^{b}{}_{\nu}\,\eta_{ab}\,,
\end{align}
where $\eta_{ab}=\mathrm{diag}(-1,1,1,1)$ is the Minkowski metric.  

To construct the static basis, we choose $e_{\hat{0}}$ to  coincide with the experimenter four-velocity (i.e., $e_{\hat{0}}$ points along the observer time direction). By considering  the static and  spherically symmetric geometry \eqref{eq:metric_static},   this procedure yields the tetrad field 
\begin{align}
e^{a}{}_{\mu} =
\begin{cases}
\mathrm{diag}\left( \sqrt{B},\, \sqrt{A},\, r,\, r\sin\theta \right), 
\quad  \;\;\; \;\;\,\, (B > 0,\; A > 0), & \\
\mathrm{diag}\left( \sqrt{-B},\, \sqrt{-A},\, r,\, r\sin\theta \right),  \;\;\;\;\;  (B < 0,\; A < 0), &
\end{cases}
\label{eq:orthonormal_frame}
\end{align}
where the signs of $B$ and $A$ ensure that the metric  retains always the Lorentzian signature. Specifically, when both $B$ and $A$ become negative (as occurs for instance inside the horizon of the Schwarzschild black hole), the coordinate basis vector $\partial/\partial r$ becomes timelike while $\partial/\partial t$ spacelike, indicating that no observer can remain at rest in  the ($r,\,\theta,\,\phi$) coordinates.

In the static frame, the experimenter  measures matter properties and expresses   the energy-momentum tensor in the following diagonal form:
\begin{align}\label{eq:diagT}
    T_{ab}= T_{\mu\nu} e_{a}{}^\mu e_{b}{}^\nu= {\rm diag} (\rho,p_1,p_2,p_3)\,,
\end{align}
where $\rho$ denotes the total  density of mass-energy, $p_1$  the radial pressure,  while $p_2$ and $p_3$ the tangential pressures \cite{Morris-Thorne1988,Rezzolla-book2013}. In this way,   the standard energy conditions translate into the following algebraic inequalities~\cite{Poisson2009}:
\begin{subequations}\label{eq:energycondition_rhoP}
   \begin{align}
    \text{NEC} &\Leftrightarrow \rho+p_i\geq0\,, \label{NEC-def}\\
     \text{WEC} &\Leftrightarrow \rho\geq 0, \;\rho+p_i\geq0\,, \label{WEC-def}\\
      \text{SEC} &\Leftrightarrow \rho+\sum_i p_i\geq0\,,\rho+p_i\geq0\,, \label{SEC-def}\\
      \text{DEC} &\Leftrightarrow \rho\geq 0, \;\rho\geq|p_i|\,, \label{DEC-def}
\end{align} 
\end{subequations}
for all $i=1,\,2,\,3$. 

It should be noted that  the  diagonalization procedure \eqref{eq:diagT} may not, in general, be extended to the location of a (Killing) horizon located at  $r=r_h$, where $B(r_h)=0$ and $e_{a}{}^{\mu}$ blows up.   Nevertheless, as proved in (Proposition 3 of) Ref.~\cite{Maeda:2021ukk},   energy criteria on the horizon can be examined by taking the limit $r \to r_h$ of relations \eqref{eq:energycondition_rhoP}, which are  defined for $B \neq 0$;  see also Ref.~\cite{Maeda:2021jdc} for a related discussion.

In the case with   $B>0$ and  $A>0$, Einstein field equations imply that the energy density and principal pressures entering Eq. \eqref{eq:diagT} are given by
\begin{subequations}\label{eq:rhoPoutBH}
    \begin{align}
    \rho &= \frac{1-(r/A)'}{8\pi r^2}\,,\\
    p_1 &= \frac{-f_0}{8\pi r^2}\,,\\
    p_2&=p_3=\frac{f_1}{8\pi r^2}\,,
\end{align}
\end{subequations}
while for  $B$ and $A$ both negative, we find
\begin{subequations}\label{eq:rhoPinBH}
\begin{align}
\rho &= \frac{f_0}{8\pi r^2}\,,\\
p_1 &= -\frac{1-(r/A)'}{8\pi r^2}\,,\\
p_2 &=p_3=\frac{f_1}{8\pi r^2}\,,
\end{align}
\end{subequations}
where the prime stands for the derivative with respect to the radial coordinate $r$ and 
\begin{align}\label{eq:deff0}
f_0(r):= &{1}-\left(\frac{1}{ A}+\frac{r B'}{ AB}\right) \,, \\
f_1(r):= &-\frac{rf_0'}{2}-\frac{r^2B'}{4}\left(\frac{1}{AB}\right)' \,.\label{eq:deff1}
\end{align}
It follows from the above   expressions that NEC \eqref{NEC-def} demands, in particular,
\begin{align}
\rho+p_1\geq0\,,\label{NEC-nec-suff}
\end{align}
thereby yielding
\begin{subequations}\label{eq:r+Pr}
\begin{align}
\frac{(AB)'}{ rA^2 B}\geq 0, \quad  \quad \text{for  $B>0$, $A>0$}\,,\\
-\frac{(AB)'}{ rA^2 B} \geq 0, \quad  \quad \text{for  $B<0$, $A<0$}\,,
\end{align} 
\end{subequations}
which, in turn, leads to  \footnote{It is worth noticing that by requiring the WEC for a regular black hole, Dymnikova \cite{Dymnikova:2001fb} obtained a result similar to Eqs. \eqref{eq:r+Pr} and~\eqref{eq:app_nec1}. }
\begin{align}\label{eq:app_nec1}
    (AB)'\geq 0\,.
\end{align}
This  inequality provides a \emph{necessary but not sufficient} condition for the NEC to hold, in the sense that if  $\left(AB \right)' <0 $ for some range of $r$, then NEC is violated throughout that region.  Relation \eqref{eq:app_nec1} constrains the function $A(r)B(r)$ to be non-decreasing. Specifically, since $AB$ is not necessarily continuous for $r>0$ (for example,  $A$ diverges at the horizon of the Schwarzschild geometry), this monotonicity is required in each separate domain of continuity.

Our subsequent investigation  will distinguish  two cases, depending on  whether $AB \neq \mathscr{C}$  (see Sec. \ref{Sec:A-B-neq-const}) or $AB = \mathscr{C}$ (see Sec. \ref{Sec:A-one-over-B}), where $\mathscr{C}$ is a  constant. The key point  is that  in the latter situation one can always choose  $A=1/{B}$ by absorbing $\mathscr{C}$ into the definition of the time coordinate.  This class of solutions is particularly relevant, as metrics like the  Schwarzschild and Reissner-Nordstr{\"o}m   fall into this category. Furthermore, we will be mainly interested in  asymptotically flat spacetimes, where for large $r$ the  components $ A(r) $ and $B(r)$ comply with
\begin{subequations}\label{eq:asy_flat}
\begin{align}
\lim_{r \to \infty} A&= 1\,,\\
\lim_{r \to \infty} B& = 1\,,
\end{align}
\text{which also imply}
\begin{align}
\lim_{r \to \infty} AB& = 1.
\end{align}
\end{subequations}

These  prescriptions thus require  the metric to approach the Minkowski form and all curvature invariants to vanish at spatial infinity, but they do not ensure a leading $1/r$  falloff. As a consequence, it is not guaranteed that the associated Misner-Sharp-Hernandez mass (and hence the ADM mass) attains a finite, nonzero value in the limit $r \to \infty$ \cite{Wald,Faraoni2021}. We will return to this point in  Sec. \ref{Sec:junction-condition}.

\subsection{Metrics with nonconstant $AB $ }\label{Sec:A-B-neq-const}

In this section, we consider metrics with  $AB\neq \mathscr{C}$, and we further divide our analysis into two scenarios:  $AB$  vanishes for some value of $r$ (see Sec. \ref{subsec:AB vanish}); $AB \neq 0 \; \forall r>0$ (see Sec. \ref{Sec:AB-neq-0}).

\subsubsection{Case with $AB $ vanishing  at some radius} \label{subsec:AB vanish}

The relation $AB=0$ defines a hypersurface where the metric is degenerate, i.e., $\det g_{\mu\nu}  = 0$. In such circumstances, index raising becomes ambiguous and the Levi-Civita connection typically diverges  (see Ref. \cite{Gunther2017} for details and e.g. Refs. \cite{Horowitz1990,Kaul2016,Kaul2017,Klinkhamer2019a,Battista2020defect,Wang2021a,Capozziello:2024ucm,Garnier2025,Capozziello:2025wwl,DeBianchi2025,Battista2026a} for some applications of degenerate metrics in cosmology and black hole models). In the following, we thus examine the limiting regime $AB \to 0$, while maintaining the validity of Einstein theory of gravity. Our aim is to identify nonstandard physical phenomena emerging in this critical setting, where conventional spacetime descriptions may require careful assessment.

The condition  $AB=0$ at some radius may arise in two distinct ways \footnote{We leave aside the  case where both $A$ and $B$ vanish simultaneously, as it yields rather exotic situations; for instance,  the energy density inevitably blows up in both the static basis and the freely falling frame, which we  introduce below.}: (i) $A=0$; (ii)  $B=0$.

In the  scenario (i), it is evident from Eqs. \eqref{eq:rhoPoutBH} and \eqref{eq:rhoPinBH} that the energy density $\rho$  blows up at  points where $A=0$. Although this pathological behavior could be avoided by  assuming $A\neq 0$,   we refrain from doing so  in order to retain generality in the subsequent analysis.  

The  situation (ii) is more subtle, as the identity $B=0$  marks the location of the event horizon(s). In this case,  the construction  \eqref{eq:orthonormal_frame} breaks down because the static experimenter four-velocity becomes infinite, and hence   the ensuing interpretation of the energy density and pressures becomes ill-defined. This  issue  can be resolved by  adopting  the orthonormal frame, say $\{e_{a'}{}^{\mu}\}$, adapted to a freely falling radial observer with  four-velocity  $v^\mu$.  

The timelike basis vector  $e_{\hat{0}'}{}^{\mu}$ is identified with $v^\mu$,  which in the geometry \eqref{eq:metric_static} is given by
\begin{align}\label{eq:4-v}
    v^\mu =\frac{\dd x^\mu}{\dd \lambda}= \left( \frac{E}{B},\ -\sqrt{\frac{1}{A}\left( \frac{E^2}{B} - 1 \right)},\ 0,\ 0 \right)\,,
\end{align}
where  we have exploited the normalization condition $ v^\mu v_\mu = -1 $,  
\begin{align}\label{eq:conservedE}
    E \equiv B(r)\,\frac{\mathrm{d}t}{\mathrm{d}\lambda}\,,
\end{align}
is the conserved energy per unit rest mass associated with the Killing vector field $ \partial_t $, and $\lambda$ represents the affine parameter. The  tetrad can now be  completed with the three mutually orthogonal spacelike  unit vectors: 
\begin{align}
e_{\hat{1}'}{}^{\mu} &= \left( -\frac{\sqrt{E^2 - B}}{B},\ \frac{E}{\sqrt{AB}},\ 0,\ 0 \right)\,, \\
e_{\hat{2}'}{}^{\mu} &= \left( 0,\ 0,\ \frac{1}{r},\ 0 \right)\,, \\
e_{\hat{3}'}{}^{\mu} &= \left( 0,\ 0,\ 0,\ \frac{1}{r\sin\theta} \right)\,,
\end{align}
where $e_{\hat{1}'}{}^{\mu}$ is constructed  to be orthogonal to $e_{\hat{0}'}{}^{\mu}$ and normalized as $ g_{\mu\nu}e_{\hat{1}'}{}^{\mu}e_{\hat{1}'}{}^{\nu}= 1 $.

The freely falling basis  allows for a  physically meaningful interpretation of the energy-momentum tensor $ T_{a'b'} = T_{\mu\nu} e_{a'}{}^\mu e_{b'}{}^\nu $ at the event horizon.  The  nonvanishing components read as 
\begin{subequations}\label{eq:rhoP_ff} 
    \begin{align}\label{eq:rho_ff}
      T_{\hat{0}'\hat{0}'} &\equiv  \rho^{\rm ff}=\frac{1}{8\pi r^2}\left[f_0 - {E^2 r}\left(\frac{1}{AB}\right)' \right] \,, \\
    T_{\hat{1}'\hat{1'}} &\equiv    p^{\rm ff}_1 =\frac{1}{8\pi r^2}\left[-1+\left(\frac{r}{A}\right)' - {E^2 r}\left(\frac{1}{AB}\right)' \right] \,,\\
     T_{\hat{2}'\hat{2}'}  &\equiv p^{\rm ff}_2= \frac{f_1}{8\pi r^2}  =  T_{\hat{3}'\hat{3}'} \equiv p^{\rm ff}_3\,, \\
         T_{\hat{0}'\hat{1}'}&= T_{\hat{1}'\hat{0}'}=\frac{E \sqrt{E^2-B}}{8\pi r^2B}\left[f_0-  1+\left(\frac{r}{A}\right)'\right]\,,  
    \end{align}
\end{subequations}
where the off-diagonal term    $T_{\hat{0}'\hat{1}'}$, which is absent in Eq. \eqref{eq:diagT} and vanishes identically in the special case  $A=1/B$,  quantifies the radial energy flux density in the experimenter local rest frame. Moreover, since the divergence stemming from the derivative of $1/(AB)$  cannot, in general, be canceled  for all possible values of the  energy   $E$,   both $\rho^{\rm ff}$ and  $ p^{\rm ff}_1$  typically become unbounded as  $AB \to 0$. 

Interestingly, in  the static tetrad \eqref{eq:orthonormal_frame} these divergences may be obscured by coordinate-dependent artifacts.  This point can be illustrated by two explicit examples where we assume that  the metric \eqref{eq:metric_static} can be expanded near a horizon located at $r=r_h$  with  the following leading-order terms:
\begin{align}
& \text{Example 1:} \qquad \qquad     A(r) \sim (r - r_h)^{-2}\,, \quad B(r) \sim (r - r_h)^3\,, \label{eq:ex1}
\\
& \text{Example 2:}  \qquad \qquad    A(r) \sim (r - r_h)^{-3}\,, \quad B(r) \sim (r - r_h)^5\,. \label{eq:ex2} 
\end{align}
As one can easily verify from Eqs. \eqref{eq:rhoPoutBH} and \eqref{eq:rhoPinBH}, the energy density and principal pressures detected by the static observer  remain finite  as $ r \to r_h $, while the corresponding quantities evaluated in the freely falling reference system  blow up. Nevertheless, no curvature singularity is  present at the horizon, since for both examples the Ricci scalar  stays regular as $ r \to r_h $
\begin{align}
&\text{Example 1:}  \qquad \qquad          R \sim \frac{-33 r^2+28 r r_h-4 r_h^2+4}{2 r^2},
\\
&\text{Example 2:}   \qquad \qquad   R \sim  \frac{-33 r^3+53 r^2 r_h-22 r r_h^2+2 r_h ^3+2}{r^2},
\end{align}
and the Kretschmann invariant $ \mathcal{K} = R_{\mu\nu\rho\sigma} R^{\mu\nu\rho\sigma} $ displays the same trend.  

In addition, the above scenarios allow us to prove that when $AB=0$ the metric can admit two   possible  behaviors: either its signature shifts from Lorentzian to Euclidean  across $r=r_h$ (as in the first case~\eqref{eq:ex1}), or  it remains Lorentzian throughout, but NEC is violated in the region $ r < r_h $ (as one can readily demonstrate in the second case \eqref{eq:ex2}).

To further explore the physical implications of the condition $AB=0$, we now  consider the radial geodesic equation. In  the spacetime \eqref{eq:metric_static}, it is given by
\begin{align}
   \frac{1}{2} \left( \frac{\dd r}{\dd \lambda} \right)^2 -\frac{E^2}{2AB}+\frac{\alpha}{2A}=0\,,
\end{align}
which resembles the dynamics of a particle with unit effective mass and zero energy moving in the
one-dimensional \qm{radial} effective potential 
\begin{align}\label{eq:veff0}
    \mathcal{V}_{\rm eff}^{\text{r}}= -\frac{E^2}{2AB}+\frac{\alpha}{2A}\,,
\end{align}
with $\alpha=1,0$ for  timelike and null geodesics, respectively. 
As $AB \to 0$, $\mathcal{V}_{\rm eff}^{\text{r}}$ tends to infinity, and we can conceive two situations: 
\begin{itemize}
    \item $AB \to 0^\pm$ when $r\to r_h^\pm$. Under these hypotheses,  we have an infinite potential well at $r_h^+$ and an infinite potential barrier at $r_h^-$. In the outer domain $r>r_h$, particles are attracted toward the horizon but cannot  cross it, while in the inner region a turning point develops before the horizon is reached.
    \item $AB \to 0^+$ for $r\to r_h^\pm$.  In this case, an infinite potential well forms on both sides of  $r=r_h$. 
\end{itemize}
In both scenarios, the horizon thus acts effectively as a two-way impenetrable surface, which  cannot be traversed  from either direction.  

This \qm{impenetrable boundary} conclusion is reinforced by the expansion scalar $\vartheta$ of the congruence of ingoing radial geodesics, which  in any static, spherically symmetric spacetime \eqref{eq:metric_static}  satisfies the  geometric identity \cite{Poisson2009,Wang:2022ews} 
\begin{align}\label{eq:expansion}
    \vartheta =\frac{\alpha (r B'+4 B)-4 E^2}{2 r \sqrt{A B} \sqrt{E^2-\alpha B}}\,.
\end{align}
Both for the null and timelike cases,  $ \vartheta \to -\infty $ as $ AB \to 0^+ $ and $E$ takes sufficiently large values.  This divergence signals  the occurrence of a caustic and may indicate the presence of conjugate points along the  congruence \footnote{In general, the existence of conjugate points signals the presence of extrema-length curves. Consider a spacelike hypersurface $ \Sigma $ and a timelike geodesic congruence orthogonal to $ \Sigma $; a point $ p $ on one geodesic of the congruence  is said to be conjugate to $ \Sigma $ if and only if the expansion scalar $ \vartheta $ tends to $ -\infty $ at $ p $. For instance, in Schwarzschild spacetime, conjugate points appear along ingoing radial timelike geodesics as $ \vartheta \to -\infty $ when $ r \to 0 $. This behavior implies that the proper lengths of these geodesics are bounded from above  in the future direction, indicating geodesic incompleteness.}, a standard ingredient in singularity-theorem arguments leading to geodesic incompleteness \cite{Wald}. It should be noted, however, that a caustic by itself does not necessarily imply a physical singularity or geodesic incompleteness. 
For example, a radial ingoing geodesic congruence in flat spacetime ($A=B=1$) also develops a caustic at the regular origin $r=0$. A related situation occurs in static regular black hole geometries, where a Cauchy horizon may be encountered before the caustic is reached by outgoing  geodesics~\cite{Carballo-Rubio:2019fnb}, but this feature alone is not interpreted as a local physical singularity.   In the present setting, however, this behavior accompanies the impenetrable-boundary structure arising as $AB\to 0$, together with the divergent stress-energy components measured in a freely falling frame.  It therefore provides an additional geometrical indication that the $AB=0$ limit represents a nonstandard and potentially pathological regime.  Interestingly, for timelike (i.e., $\alpha=1$) infalling radial paths,  $\vartheta$  can become positive when   $r B'  > 4(E^{2}-B)$, and    energy conditions may be violated. We will discuss this situation in detail in Sec.~\ref{sec:wec}.

From the above analysis, it is clear that when the metric retains the Lorentzian signature, the fulfillment of the NEC implies that a singularity may arise at  points  where $AB \to 0^+$, even though curvature  may not attain unboundedly large values.  This outcome is fully consistent with the expectations of classical singularity theorems \cite{Hawking:1970zqf}. Although our examination is primarily based on general relativity and focuses on the region $r>0$ of the spacetime, 
the result that $AB \to 0$ implies a potential singularity can be straightforwardly generalized 
to the case $r=0$ or to other modified theories of gravity admitting static, spherically symmetric metrics. This conclusion is supported by several models with $A\neq 1/B$,  such as the Janis-Newman-Winicour spacetime~\cite{Joshi:2011zm,Joshi:2013dva,Shaikh:2018lcc} 
and certain Brans-Dicke solutions~\cite{HassanPuttasiddappa:2025tji,Bergh:1980mta}, 
in which naked singularities appear at $r=0$ accompanied by the vanishing of the product $AB$.

The examination of this section can be relevant for numerous effective or quantum-inspired black holes, in which the underlying corrections lead to scenarios where $ A(r) $ remains finite as $ B(r) \to 0 $ (see e.g. Refs. \cite{DelPiano2023,Battista2023,Wang-Battista2025,Wei2025,Hohenegger2025}), so that the product $ AB \to 0 $ at the horizon. As we have shown before, in this setup new effects departing from the classical Schwarzschild paradigm can arise. Such phenomena may indicate a breakdown of the equivalence principle at the horizon and suggest the presence of novel near-horizon physics beyond the classical pattern~\cite{Chapline:2000en}. 

Similar effects as those encountered in this section can emerge in non-static and  spherically symmetric geometries, for which we investigate the NEC in Appendix \ref{Sec:Appendix}. 

\subsubsection{Case with $AB$ different from zero }\label{Sec:AB-neq-0}

According to the discussion of the previous section, the requirement  $AB \neq 0$ serves to prevent several undesired features, including NEC violations and metric-signature changes. For this reason,  we now assume that $AB \neq 0$  throughout the spacetime except at $r = 0$, where a curvature singularity may exist.

The necessary condition~\eqref{eq:app_nec1} for the NEC fulfillment implies that, under the hypothesis that  $AB$ is a smooth function, $(AB)'$ can only be zero at isolated points. Therefore, $AB$ is monotonically increasing in its domains except  where $(AB)'=0$. Since, in an asymptotically flat spacetime, the metric functions $ A(r) $ and $B(r)$ comply with Eq.~\eqref{eq:asy_flat},  
the quantity $AB$ is  restricted to values not exceeding  $1$. These considerations lead to the following conclusion:
\begin{quote}
In the asymptotically flat, static, and spherically symmetric spacetime \eqref{eq:metric_static}, the smooth function $AB$ must satisfy one of the following conditions, otherwise either NEC is violated or the metric signature changes from Lorentzian to Euclidean--both of which can potentially lead to geodesic incompleteness:
\begin{itemize}
\item[(i)]  $AB$ is constant for all $r>0$\,,
\item[(ii)]  $0<AB\leq 1$ for all $r>0$.
\end{itemize}
\end{quote}
In this situation, the  metric can be expressed as 
\begin{align}\label{eq:metric_static2}
{\rm d}s^2= - B(r) {\rm d}t^2  + \frac{F(r)}{B(r)} {\rm d}r^2 + r^2 {\rm d}\Omega^2\,,
\end{align}
with 
\begin{align}
    0< F(r)\leq 1\,.
\end{align}

Notice that, for a black hole solution,  the above statement  entails that a constant-$r$ null hypersurface must also be a metric horizon. Indeed, the latter  is identified by the equation $B(r_h)=0$, and hence the constraint that $AB$ remains finite yields $A\to \infty$ at $r=r_h$.  

\subsection{Solutions with $A(r)=1/B(r)$}\label{Sec:A-one-over-B}

We now turn to the case $AB=\mathscr{C}$, where  the constant  $\mathscr{C}$ is nonzero in order to avoid a  trivial,  nonphysical solution. 

As pointed out before, in this setup it is   always possible to set $A=1/B$  and  then express the metric  as
\begin{align}\label{eq:metric_static-3}
{\rm d}s^2= - B(r) {\rm d}t^2  + \frac{{\rm d}r^2}{B(r)}  + r^2 {\rm d}\Omega^2\,.
\end{align}

This spacetime belongs to the so-called Kerr-Schild class (see e.g. Refs. \cite{Jacobson2007,Ovalle2023,Casadio2024}). We now examine the constraints imposed by the NEC (Sec. \ref{Sec:NEC-constraints-AB-1}), WEC (Sec. \ref{Sec:WEC-constraints-AB-1}),  SEC (Sec.  \ref{Sec:SEC-constraints-AB-1}), and DEC (Sec. \ref{Sec:DEC-constraints-AB-1}). As we will see,  our investigation allows us to devise an algorithm  for generating   solutions to Einstein field equations of the form \eqref{eq:metric_static-3} that automatically comply with NEC (see Eqs. \eqref{y-condition} and \eqref{eq:def_y} below).

\subsubsection{Constraints from NEC}\label{Sec:NEC-constraints-AB-1}

By adopting the static tetrad (cf. Eq. \eqref{eq:orthonormal_frame}), we find  from Eq. \eqref{eq:rhoPoutBH} that the energy density and principal pressures for the spacetime \eqref{eq:metric_static-3} take the form
\begin{subequations}\label{eq:rhoP}
    \begin{align}\label{eq:rhoP-a}
    \rho &= \frac{1-B-r B'}{8\pi r^2}\,,\\
    p_1 &= -\rho\,,\\
    p_2&=p_3=\frac{rB''+2B'}{16\pi r}\,.
\end{align}
\end{subequations}
This means that the NEC \eqref{NEC-def} is satisfied if and only if 
\begin{align}
y(r)\geq 0,    \label{y-condition}
\end{align}
where we have introduced the $r$-dependent function
\begin{align}\label{eq:def_y}
y(r)= r^2 B''(r)-2 B(r) +2\,.
\end{align} 
It is worth  noticing that Eqs. \eqref{y-condition} and  \eqref{eq:def_y} are equivalent to the null converge condition $R_{\mu \nu} k^\mu k^\nu \geq 0$ (where recall $k^\mu$ is a generic null four-vector; see Eqs. (9) and (10) in Ref. \cite{Borissova2025b}), and thus provide a clear illustration of how this geometric relation corresponds to NEC in Einstein gravity.

To explore some explicit examples meeting the constraint \eqref{y-condition}, we consider the \emph{ansatz} 
\begin{align}
y(r)=c_nr^n,\label{ansatz-1}
\end{align}
where $c_n \geq 0$ are constants and  $n$ is  an integer (this choice is made for  simplicity, as our results extend straightforwardly also to non-integer values of $n$). In this way, we   isolate the elementary power-law building blocks of non-negative functions $y(r)$. 
Owing to the linearity of Eq.~\eqref{eq:def_y}, more general NEC-compatible profiles can subsequently be generated by superposition, as will be shown shortly. In view of  Eq. \eqref{ansatz-1},  Eq. \eqref{eq:def_y} can be read as an ordinary differential equation whose solutions produce metric functions $B(r)$  that automatically obey the NEC. Let us consider the following choices for $y(r)$:
\begin{itemize}
\item $y(r) =0$. In this case, Eq.~\eqref{eq:def_y}  becomes a homogeneous differential equation and we readily obtain  
       \begin{align}\label{eq:basic}
           B(r) =1-\frac{d_1}{r}-d_2 r^2\,,
       \end{align}
with $d_{1,2}$  (dimensionful) integration constants.   Within our class of solutions,  Eq.~\eqref{eq:basic} is the only one that gives rise to an isotropic energy-momentum tensor,  whereas the other configurations necessarily involve anisotropic pressures (cf. Eq.~\eqref{eq:rhoP}). Setting $d_{1}=\RS\equiv 2M$ and $d_2>0$, we obtain the Schwarzschild-de Sitter metric. Thus, to maintain consistency with this physical interpretation, we will henceforth fix $d_{1}=\RS\equiv 2M$ (more details on the parameter $M$ will be given in Sec. \ref{Sec:junction-condition}).
\item $y(r) =c_0$. For this scenario, Eq. \eqref{eq:def_y} yields  the solution
       \begin{align}\label{eq:c0constant}
           B(r) =1-\frac{\RS}{r}-d_2 r^2-\frac{c_0}{2}\,.
       \end{align}
\item $y(r)={c_{-1}}/{r}$\,. 
\begin{align}\label{new_metric_log}
    B(r)=1-\frac{\RS}{r}-d_2 r^2-\frac{c_{-1} \ln r}{3r}\,.
\end{align}
\item $y(r)={c_2}{r^2}$\,.
\begin{align}
    B(r)=1-\frac{\RS}{r}-d_2 r^2+\frac{c_2 r^2 \ln r}{3}\,.
\end{align}
\item $y(r)={c_m}{r^m}$ with $m\neq 2 $ and $m\neq -1$\,. 
\begin{align}\label{eq:c0r_2}
    B(r)=1-\frac{\RS}{r}-d_2 r^2+\frac{c_m r^m }{m^2-m-2}\,.
\end{align}
In this situation, we can recover  the Reissner-Nordstr{\"o}m metric by taking $m=-2$ and $d_2=0$. 
\end{itemize}

In the above discussion,  we have neglected  dimensions for simplicity. Since the function $ y(r)$ is dimensionless, it is convenient to take $c_n$ to be dimensionless as well. To ensure this,  a length scale $r_d$ should be introduced into  Eq. \eqref{ansatz-1}, which can be thus reexpressed as   $y(r)=c_n\left(r/r_d\right)^n$. For instance,  setting  $y=c_{-1}(r_d/r)$ Eq.~\eqref{new_metric_log}  reads as    (recall that $d_2$ is dimensionful)
\begin{align}
B(r)=1-\frac{\RS}{r}-d_2 r^2-\frac{c_{-1} r_d \ln (r/r_d)}{3r}\,.
\label{B-r-with-length-scale}    
\end{align}
The parameter $r_d$ could be  identified with $\RS$ or  another relevant physical scale, such as the Planck length, depending on the  context of the problem. In the remainder of this section, we will omit $r_d$ for simplicity, but we will restore  it   in the examination of Sec. \ref{Sec:geometry-log-correction}.

Since relation ~\eqref{eq:def_y} gives rise to a linear differential equation, the superposition principle applies,   guaranteeing  that any linear combination of Eqs. \eqref{eq:c0constant}--\eqref{eq:c0r_2} still solves Eq. \eqref{eq:def_y}. Consequently, if we write 
\begin{align}\label{eq:assumption_y}
    y(r)= \sum_{n} c_n r^n \,, 
\end{align}
then  $B(r)$ takes the general form 
\begin{align}\label{eq:general_Bslo}
B(r)=&1-\frac{\RS}{r}-d_2 r^2-\frac{c_{-1} \ln r}{3r}+\frac{c_2 r^2 \ln r}{3} +\sum_{ n\neq-1,2}\frac{c_n r^n }{n^2-n-2}\,,
\end{align}
and  condition  \eqref{y-condition} automatically holds as long as $c_n\geq 0$ for all terms in  the sum \eqref{eq:assumption_y}.

Since we have adopted an \emph{ansatz} to enforce the requirement \eqref{y-condition}, there also exist  NEC-satisfying configurations  that are not captured by Eq.~\eqref{eq:general_Bslo}, such as the well-known regular black hole models (which, we recall, typically spoil the SEC)~\cite{Dymnikova2003,Bonanno2000,Hayward2005}. Nevertheless, our result~\eqref{eq:general_Bslo} provides a straightforward criterion for assessing NEC violations in numerous modified Schwarzschild and quantum-corrected black hole geometries studied in the literature (see e.g. Refs. \cite{Bargueno2016,Lewandowski:2022zce,DelPiano2023,Hohenegger2025,Ahmed2026b}). For example, if $B(r)$ has the form
\begin{align}\label{B-r-function-example-1}
    B(r)=1-\frac{\RS}{r}+\frac{c_i}{r^i}\,, \;\;\text{with}\; i\geq 2\,,
\end{align}
NEC is  respected if and only if $c_i\geq 0$, and similarly  for 
\begin{align}\label{B-r-function-example-2}
   B(r)=  1- \frac{\RS}{r}+ \frac{c_i}{r^i}+ \frac{c_m}{r^m}\,, \;\;\text{with}\;2\leq i<m \, , 
\end{align}
NEC is  met if and only if $c_i\geq 0$ and $c_m\geq 0$. 

We  emphasize that the lower bound $c_n\geq 0$ for  the coefficients  in the sum  \eqref{eq:assumption_y}  represents in general a sufficient, though not a necessary, condition for  the NEC to hold (the cases \eqref{B-r-function-example-1} and \eqref{B-r-function-example-2} admit necessary \emph{and} sufficient conditions  because they contain few parameters). To illustrate this point, we now discuss a scenario where some $c_n$ may be negative while the constraint \eqref{y-condition}, and hence the NEC, is fulfilled.

For physical relevance, we  consider an asymptotically flat spacetime, with
\begin{align}\label{eq:Br_c_imn}
   B(r)=  1- \frac{\RS}{r}+ \frac{c_i}{r^i}+ \frac{c_m}{r^m}+ \frac{c_n}{r^n}\,, \quad \text{where}\; 2\leq i <m<n.
\end{align}
The function $y(r)$ defined in Eq.~\eqref{eq:def_y} then becomes 
\begin{align}
y(r)&=\frac{c_i(i^2+i-2)}{r^i}+\frac{c_m(m^2+m-2)}{r^m}+\frac{c_n(n^2+n-2)}{r^n}\,,
\end{align}
and exhibits the asymptotic behavior  
 \begin{align}
y(r) \to 
\begin{cases} 
\frac{c_i(i^2+i-2)}{r^i}, & \text{as } r \to \infty\,, \\
\frac{c_n(n^2+n-2)}{r^n}, & \text{as } r \to 0\,.
\end{cases}
\end{align}
To secure  condition \eqref{y-condition} in the limits $r\to0 $ and $r\to \infty$, it is necessary that $c_n\geq 0$ and $c_i\geq 0$, while ensuring that  $y(r)$  remains non-negative throughout its  domain further requires that
\begin{align}\label{eq:inequality_c_m}
c_m (m^2+m-2) &\geq -c_i(i^2+i-2)r^{m-i}-c_n(n^2+n-2)/r^{n-m}\,.
\end{align}
The maximization of the right-hand side of Eq.~\eqref{eq:inequality_c_m} is given by
\begin{align}\label{eq:f1}
&-\mathcal{F}=-{c_i(i^2+i-2)}\left[ \frac{c_n(n^2+n-2)(n-m)}{c_i(i^2+i-2)(m-i)}\right]^{\frac{m-i}{n-i}}
\nonumber \\
&- {c_n(n^2+n-2)}\left[ \frac{c_i(i^2+i-2)(m-i)}{c_n(n^2+n-2)(n-m)}\right]^{\frac{n-m}{n-i}}\,,
\end{align}
which means that  the necessary and sufficient condition to satisfy the NEC is
\begin{align}
    c_{i,n}\geq 0\,, \;\; \text{and} \;\; c_{m}\geq -\frac{\mathcal{F}}{m^2+m-2}\,.
\end{align}
Since $\mathcal{F}$ is positive, it follows that $c_m$ can indeed be negative, as pointed out before. 

From our analysis it is  clear that Eqs. \eqref{y-condition} and \eqref{eq:def_y} provide a systematic method  for  producing metrics in the Kerr-Schild form \eqref{eq:metric_static-3} that adhere to the NEC, the only input being an \emph{ansatz} for the function $y(r)$, as done in Eqs. \eqref{ansatz-1} and \eqref{eq:assumption_y}.  A key advantage of this procedure is that it enables one to determine a broad family of geometries, including the well-known  Schwarzschild, Schwarzschild-de Sitter, and  Reissner-Nordstr{\"o}m solutions. Table~\ref{tab:energy_conditions1}  summarizes the necessary \emph{and} sufficient conditions for the NEC to apply for some of the ensuing configurations.
\begin{table*}[htbp!]
    \centering
    \renewcommand{\arraystretch}{2} 
    \begin{tabular}{|c|c|}
        \hline
        \textbf{Metric function $B(r)$} & \textbf{Necessary and sufficient conditions for the NEC}  \\
        \hline
        $1 - \frac{\RS}{r} + \frac{c_i}{r^i}$ & $c_i \geq 0$  \\ 
        \hline
        $1 - \frac{\RS}{r}+ \frac{c_i}{r^i} + \frac{c_m}{r^m}$ & $c_i \geq 0$, \quad  $c_m \geq 0$  \\ 
        \hline
        $1 - \frac{\RS}{r} + \frac{c_i}{r^i} + \frac{c_m}{r^m} + \frac{c_n}{r^n}$ & $c_i \geq 0$, \quad  $c_n \geq 0$, \quad  $c_m \geq -\mathcal{F}/({m^2+m-2})$  \\ 
        \hline
    \end{tabular}
    \caption{Necessary and sufficient conditions for the NEC to be respected by static and spherically symmetric metrics of the Kerr-Schild form \eqref{eq:metric_static-3}. For all the solutions analyzed, it is understood that $2 \leq i< m < n$.  The parameter $\mathcal{F}$ can be read off from Eq.~\eqref{eq:f1}. Each case generalizes the previous one, and accordingly the last one includes the geometries considered in the first two rows. }
    \label{tab:energy_conditions1}
\end{table*}

Since the NEC   does not necessarily guarantee compliance with the  WEC, SEC, and DEC, in the next sections  we  examine the restrictions due to these tighter conditions and  show that they lead to additional limitations on the metric function \eqref{eq:general_Bslo}.  

\subsubsection{Constraints from WEC}\label{Sec:WEC-constraints-AB-1}
\label{sec:wec}

When we impose the validity of the WEC \eqref{WEC-def}, it follows from Eq. \eqref{eq:rhoP-a} that, in addition to  Eq. \eqref{y-condition}, the metric component $B(r)$ is subject to the further  constraint 
\begin{align}
1-B-rB'& \geq 0\,,\label{eq:wec2}
\end{align}
which, for the general solution \eqref{eq:general_Bslo},   translates into 
\begin{align}\label{eq:wec_B}
9 d_2 r^2+\frac{c_{-1}}{r}-c_2 r^2(1+3 \ln r)-3\sum _{n\neq -1,2}\frac{ c_n r^n}{n-2} \geq 0 \,.
\end{align}
To assess whether the WEC holds,  we  analyze each term on the left-hand side separately. Taking $d_2\geq 0$ guarantees that the first contribution complies with the WEC, while the second  does so automatically, since the NEC already enforces $c_{-1}\geq 0$. However, the third  factor proportional to $c_2$ can become negative, and hence  satisfying the WEC necessitates setting $c_2=0$. Finally, for the remaining sum, a sufficient condition for non-negativity is $n<2$, along with  the bound  $c_n\geq 0$, as already mandated by NEC. 

We can now prove that if the expansion scalar $\vartheta$, computed in Eq.~\eqref{eq:expansion}, becomes positive along ingoing radial timelike geodesics, then WEC is necessarily spoiled. To this end, we  consider a  massive particle initially  at rest at some radius $r$, which then begins to free-fall toward $r = 0$. This onset is triggered when
\begin{align}
\frac{\dd }{\dd r} \mathcal{V}_{\rm eff}^{\rm r,timelike}>0,\label{condition-effective-pot}
\end{align}
where the \qm{timelike} effective potential $\mathcal{V}_{\rm eff}^{\rm r,timelike}$ is obtained from Eq.~\eqref{eq:veff0} by substituting $A = 1/B$ and $\alpha = 1$. The above inequality  entails 
\begin{align}
B' > 0\,, \label{wec-bound-1}    
\end{align}
while the WEC imposes Eq.~\eqref{eq:wec2}, which is equivalent to  
\begin{align}\label{eq:wec-1}
    r B' < 1 - B \, .
\end{align}
Taken together,  the bounds \eqref{wec-bound-1} and \eqref{eq:wec-1}  imply that $1 - B > 0$. Now, let us examine radially moving massive bodies released from rest in the asymptotically flat region $r \to \infty$, for which the conserved energy takes the value $E = 1$.  In this situation, the condition $\vartheta > 0$ yields  
\begin{align}
    r B' > 4 (1 - B) \,,
\end{align}
which is clearly incompatible with Eq.~\eqref{eq:wec-1} in the physically relevant domain $B < 1$.
Consequently, within this setup, if the WEC holds the congruence must satisfy $\vartheta\le 0$, corresponding to the non-expanding regime in which observers experience neither defocusing nor local repulsive effects.

\subsubsection{Constraints from SEC} \label{Sec:SEC-constraints-AB-1}

Starting from Eq. \eqref{eq:rhoP}, one readily finds that  the SEC \eqref{SEC-def} is   valid if and only if Eq. \eqref{y-condition} is supplemented by 
\begin{align}
    B''+\frac{2 B'}{r} & \geq 0\,,\label{eq:sec_B2}
\end{align}
which, when applied to the metric coefficient  \eqref{eq:general_Bslo},  reduces to 
\begin{align}\label{eq:sec_B}
-6 d_2+\frac{c_{-1}}{3 r^3}+c_2(2 \ln r+{5}/{3})+\sum _{n\neq -1,2} \frac{n c_n  r^{n-2}}{n-2} \geq 0 \,.
\end{align}
If $d_2>0$, the first term on the left-hand side  circumvents  SEC, a result  fully consistent with the interpretation of $d_2>0$ as corresponding to a positive cosmological constant. The second factor complies with the SEC as $c_{-1}\geq 0$ owing to NEC, while for $c_2>0$ the third one violates the SEC,  since it can become  negative  for sufficiently small  $r$.  A sufficient condition for the last contribution involving the sum to remain  non-negative is $  n\leq 0 \cup n> 2$ jointly with $c_n\geq 0$. 

Notice that, since our method is framed in Einstein gravity, inequalities \eqref{y-condition} and \eqref{eq:sec_B2} encode the timelike convergence condition,  with the latter constituting the  additional constraint not already enforced by the null convergence condition (see Eq. (8) in Ref. \cite{Borissova2025}).

\subsubsection{Constraints from DEC} \label{Sec:DEC-constraints-AB-1}

The DEC~\eqref{DEC-def} demands that Eq.~\eqref{y-condition} be combined with Eq.~\eqref{eq:wec2} and 
\begin{align}\label{eq:DecB2}
    r^2 B''(r)+4 r B'(r)+2 B(r)-2 \leq 0 \, ,
\end{align}
which gives
\begin{align}\label{eq:dec_B}
    12 d_2 r^2 + \frac{c_{-1}}{3 r}
    - c_2 r^2\!\left(\frac{7}{3}+4 \ln r\right)
    - \sum_{n\neq -1,2}\frac{c_n (n+2) r^n}{n-2} \geq 0 \, .
\end{align}
This inequality  holds for all $r>0$  if $d_2 \geq 0$, $c_{-1} \geq 0$, and $c_{2}=0$, which are already demanded by the WEC. Moreover, a sufficient condition for the sum to be  non-negative is $-2 \leq n < 2$, which is stronger than the requirement imposed by the WEC. 
\begin{table*}[htbp!]
    \centering
    \renewcommand{\arraystretch}{2} 
    \begin{tabular}{|c|c|}
        \hline
        \multicolumn{2}{|c|}{ \hspace{1cm} \textbf{Metric function:}  $
            B(r) = 1 - \frac{\RS}{r} - d_2 r^2 - \frac{c_{-1} \ln r}{3r} + \frac{c_2 r^2 \ln r}{3} + \sum\limits_{{n \neq -1, 2}}  \frac{c_n r^n}{n^2 - n - 2}
        \hspace{1cm} $  } \\ 
        \hline
        \textbf{Energy condition} & \textbf{Sufficient conditions on the coefficients of $B(r)$} \\ 
        \hline
        NEC & $c_{-1} \geq 0$, \; $c_2 \geq 0$, \; $c_n \geq 0$ \\ 
        \hline
        WEC & $d_2 \geq 0$, \; $c_{-1} \geq 0$, \; $c_2 = 0$, \; $n < 2$, \; $c_n \geq 0$ \\ 
        \hline
        SEC & $d_2 \leq 0$, \; $c_{-1} \geq 0$, \; $c_2 = 0$,   \; $  n\leq 0 \cup n> 2$, \; $c_n \geq 0$ \\ 
        \hline
        DEC & $d_2 \geq 0$, \; $c_{-1} \geq 0$, \; $c_2 = 0$, \; $-2 \leq n < 2$, \; $c_n \geq 0$ \\
        \hline
    \end{tabular}
    \caption{Sufficient conditions on the coefficients of the metric function  $B(r)$ for the classical energy criteria to be satisfied.  Restrictions become increasingly stringent for stronger conditions. All four energy criteria are simultaneously fulfilled only in the very restricted setup with $d_2=0$, $c_{-1} \geq 0$, $c_2=0$, and  the sum running over $-2\leq  n\leq 0$ with  $c_n \geq 0$.  }
    \label{tab:energy_conditions2}
\end{table*}

The constraints imposed on the general metric function~\eqref{eq:general_Bslo}
by the classical energy conditions are summarized in Table~\ref{tab:energy_conditions2}. As expected, the SEC implies the NEC but not the WEC, while the DEC implies both the WEC and the NEC. Moreover, progressively stronger energy conditions lead to increasingly restrictive limitations on the metric.

\section{Logarithmically corrected spacetime}\label{Sec:geometry-log-correction}

In the previous section, we have set up a  procedure for deriving NEC-fulfilling static and spherically symmetric metrics with $A=1/B$, starting from Eqs. \eqref{y-condition} and \eqref{eq:def_y}. Within the family of resulting geometries, we can single out a particularly relevant case by setting $d_2 = 0$ in Eq.~\eqref{new_metric_log} (see also Eq. \eqref{B-r-with-length-scale}). In this way, we obtain the asymptotically flat spacetime featuring a logarithmic correction to the Schwarzschild solution
\begin{align}\label{eq:metric_log}
{\rm d}s^2=& - \left[1-\frac{\RS}{r}-\frac{c_{-1} r_d\ln (r/r_d)}{3r}\right]{\rm d}t^2 
+\left[1-\frac{\RS}{r}-\frac{c_{-1} r_d \ln (r/r_d)}{3r}\right]^{-1}{\rm d}r^2 + r^2 {\rm d}\Omega^2\,,
\end{align}
where we have restored the length scale $r_d$ for the upcoming discussion. By virtue of Einstein equations, this metric can be associated with the energy-momentum tensor
\begin{align}\label{EMT_log}
   \tensor{T} {_\mu ^\nu}&=\frac{c_{-1} r_d }{24\pi r^3}\,{\rm diag}\left(-1,\,-1,\,\frac{1}{2},\,\frac{1}{2} \right)\,,
\end{align}
which displays an anisotropic pressure distribution. Remarkably, one can verify that  \emph{all} energy criteria \eqref{eq:energycondition_rhoP} are satisfied provided that $c_{-1}$ is non-negative.

Interestingly, a line element of the form \eqref{eq:metric_log} also appears in Ref.~\cite{Li:2012zx}, where it is used to study  a phantom field as a  dark-matter candidate in galactic environments, although the corresponding logarithmic coefficient (related to the parameter $a$ in Ref.~\cite{Li:2012zx}, up to normalization and sign conventions) is left arbitrary there. Instead, our results furnish an additional theoretical restriction on the model by fixing  the sign of $c_{-1}$. This highlights that our algorithm based on Eqs. \eqref{y-condition} and \eqref{eq:def_y} has a genuine physical foundation, as it permits the generation of metrics that arise in phenomenologically motivated scenarios while, at the same time, constraining their parameter space. Moreover, the logarithmic function in Eq. \eqref{eq:metric_log} involves two independent constants, as opposed to the single term used in Ref. \cite{Li:2012zx}. As we will demonstrate below, this facet gives our framework a substantially richer structure.

In addition,  logarithmic contributions also appear, for example, in generalized  Ba\~{n}ados-Teitelboim-Zanelli (BTZ) black holes \cite{Mendez-Zavaleta2026}, in hairy black holes in Horndeski gravity \cite{Bergliaffa2021}, and in generic metric-affine  frameworks \cite{DAmbrosio2021}. Moreover, though the physical origin is different,  logarithmic  radial factors occur in  the post-Newtonian (near-zone) expansion of the (harmonic-coordinate) metric (where they encode   hereditary  tails and  tails-of-tails effects;  see Eq. (87) in  Ref. \cite{Blanchet2013-review}). This property characterizes the so-called polyhomogeneous spacetimes  in the context of radiating systems \cite{Chrusciel1993,He2026}.

In this section, we perform a detailed analysis of  Eq.~\eqref{eq:metric_log}, which we  regard  as  an effective static exterior geometry of a compact object, to be matched in principle   to a (possibly regular) interior and  intended to be valid in the domain probed by null and timelike geodesics.
We begin by exploring its key  features, such as the $r=0$ singularity (see Sec. \ref{Sec:r-0})  and the possible existence of  horizon(s) (see Sec. \ref{Sec:event-horizon}). Afterwards, we carry out a preliminary assessment of the physical viability of this model. In Sec. \ref{Sec:photon-sphere},  we discuss the emerge of photon spheres and innermost stable circular orbits (ISCOs), and compare our results with bounds provided by the Event Horizon Telescope (EHT) collaboration. In Sec. \ref{Sec:BH-mimickers}, we classify the horizonless and horizon-bearing configurations, while solar-system-like constraints are examined in  Sec. \ref{Sec:solar-system-tets}. Finally, in Sec. \ref{Sec:junction-condition} we evaluate junction conditions  required to join the log-corrected solution to an exterior Schwarzschild spacetime.

\subsection{The $r=0$ singularity}\label{Sec:r-0}

The Ricci and Kretschmann scalars characterizing the log-metric~\eqref{eq:metric_log} read as 
\begin{align}
R=&\frac{c_{-1} r_d}{3r^3}\,, 
\nonumber \\
\mathcal{K}=&\frac{1}{9 r^6}\Bigl\{4 c_{-1} r_d \ln \left(r/r_d\right) \left[3 c_{-1} r_d \ln \left(r/r_d\right)-5 c_{-1} r_d+18 \RS\right]+13 \left(c_{-1} r_d\right)^2 \nonumber \\
&-60 c_{-1}  r_d \RS +108 \RS^2\Bigr\} \,,
\end{align}
respectively. Since they both blow as $r \to 0^+$, a curvature singularity is  present at $r=0$, where also  the energy density (i.e., $-\tensor{T} {_0 ^0}$) blows up for any nonzero   $c_{-1}$ in view of Eq.~\eqref{EMT_log}.

In this regard, an intriguing feature of the spacetime  is revealed by the study of  the geodesic equation. Upon restricting to  the  equatorial plane $\theta=\pi / 2$, it takes the form 
\begin{align}\label{eq:geodesic_log}
    \frac{1}{2}\left(\frac{\dd r}{\dd \lambda}\right)^2+V_{\rm eff}=\frac{E^2}{2}\,,
\end{align}
where the conserved energy $E$ has been defined in Eq.~\eqref{eq:conservedE} and 
 \begin{align}\label{eq:veff}
   V_{\rm eff}= \frac{1}{2} \left[1-\frac{\RS}{r}-\frac{c_{-1} r_d\ln (r/r_d)}{3r}\right]\left(\frac{L^2}{r^2}+\alpha\right)\,,
\end{align}
is the effective potential, where,  for simplicity,  we have adopted a different definition with respect to  Eq.~\eqref{eq:veff0}; in addition,   
\begin{align}\label{eq:defL}
L\equiv r^2\frac{\dd \phi}{\dd \lambda}\,,
\end{align}
is the conserved angular momentum associated with rotational symmetry, and, like before, $\alpha=1$ and $\alpha=0$ correspond to timelike and null geodesics, respectively. 

Since $c_{-1}>0$ as required by the energy conditions, $ V_{\rm eff} \to +\infty$ as $r\to 0$, except in the radial null case (i.e., when $L=\alpha=0$).  This means that both timelike and non-radial null geodesics cannot reach the singularity, unlike radial null rays.

\subsection{Horizon structure}\label{Sec:event-horizon}

Since we are dealing with a static and spherically symmetric geometry, the locations of possible  horizons are determined by the condition
\begin{align}
B(r_h)=1-\frac{\RS}{r_h}-\frac{c_{-1} r_d\ln (r_h/r_d)}{3r_h}=0,
\end{align}
which yields  the solution 
\begin{align} \label{event-hor-solution}
r_h=-\frac{c_{-1}r_d}{3} W_k \left[-\frac{3}{c_{-1}} {\rm e}^{-3\RS/\left(c_{-1}r_d\right)} \right].
\end{align}
Here,  $W_k$ denotes the $k$-th branch of the (multivalued) Lambert $W$ function \cite{Corless1996,Nist},  implicitly defined by the relation
\begin{align}
W(x) {\rm e}^{W(x)}=x \,.     \label{Lambert-W-equation} 
\end{align}
When the argument $x$ is real,  only the two real-valued branches $W_0$  and $W_{-1}$ (referred to as principal and lower  branches, respectively) are relevant, and thus Eq. \eqref{Lambert-W-equation} admits the following   structure:
\begin{itemize}
\item $x \geq 0$: there exists exactly one real root, namely $W_0$; 
\item $-1/{\rm e} < x < 0$:  two   \emph{distinct} and real solutions occur, i.e.,    $W_0$ and $W_{-1}$;  
\item $x=-1/{\rm e}$: the two real branches $W_0$ and $W_{-1}$ merge into a single degenerate value, owing to the identity $W(-1/{\rm e}) = -1$;
\item $x<-1/{\rm e}$:   no real solutions are  possible.
\end{itemize}
Therefore, the presence of the Lambert $W$ function in formula \eqref{event-hor-solution} entails that the spacetime may exhibit two horizons, a single  horizon (which we can view as an extremal horizon, since the two horizons associated with $W_0$ and $W_{-1}$ coalesce), or no horizon at all. We now investigate this point in detail.

From Eq. \eqref{event-hor-solution}, we see that it is convenient to express the argument of  $W$  in terms of  the $c_{-1}$-dependent function
\begin{align}
x(c_{-1}) &= -\frac{3}{c_{-1}} \, {\rm e}^{- 3 \RS/(c_{-1} r_d )} \,,
\end{align}
which, under our assumptions (recall that  $c_{-1}$, $\RS$, and $r_d$ are positive) is always negative and  attains a minimum at
\begin{subequations}
\label{min-and-value-1}
\begin{align}
c_{-1,{\rm min}} &= \frac{3 \RS}{r_d}\,,
\end{align}
\text{where} 
\begin{align}
x_{\rm min} &= -\frac{r_d}{\RS \, \ee}.
\label{x-min-horizon}
\end{align}
\end{subequations}
As $c_{-1} \to 0^+$ or $c_{-1} \to \infty$, $x(c_{-1})$ approaches zero from below, and 
thus its overall range  is 
\begin{align}
x(c_{-1}) \in \Big[x_{\rm min}, \, 0\Big), \label{range-x-c}
\end{align}
where we  have excluded $x(c_{-1})=0$, because it would give  $r_h=0$ in Eq. \eqref{event-hor-solution} due to the fact that the Lambert  function satisfies $W(0)=0$.

Now, taking into account the properties of $W$  discussed above and the fact that  $r_d$ is \emph{a priori} a free parameter, we can distinguish  the following situations: 
\begin{itemize}
\item $r_d < \RS$. From Eq. \eqref{x-min-horizon}, one has $x_{\rm min} > - 1/\ee$. By Eq. \eqref{range-x-c}, this  implies that $x(c_{-1})>-1/\ee \; \forall c_{-1}>0$, which reveals that  the argument of the Lambert function always lies in the interval where both real branches $W_0$ and $W_{-1}$ exist. Consequently, for any $c_{-1}>0$, the spacetime~\eqref{eq:metric_log} always supports two horizons. 
\item $r_d = \RS$. In this case, Eq. \eqref{min-and-value-1} indicates  that the function $x(c_{-1})$ takes at $c_{-1,{\rm min}}=3$ the minimum  $x_{\rm min} =-1/\ee$.  Therefore, when $c_{-1} = 3$, the branches $W_0$ and $W_{-1}$ coincide, giving rise to a single (degenerate) horizon located at $r=\RS$. For $c_{-1}\neq 3$, we have $x(c_{-1})>-1/\ee$, and hence the metric develops two horizons, one of which  always  at $r=\RS$. In particular, for $c_{-1} < 3$, $r=\RS$ corresponds to the outer horizon.
\item $r_d > \RS$. Since  $x_{\rm min} < - 1/\ee$,  the metric may allow for  two, one, or  no horizon, depending on the value of $c_{-1}$. In this latter scenario, $r=0$ becomes a naked singularity.
\end{itemize}

\subsection{Photon spheres and innermost stable circular orbits}\label{Sec:photon-sphere}

\begin{figure}[htbp!]
    \centering
    \includegraphics[width=0.65\textwidth]{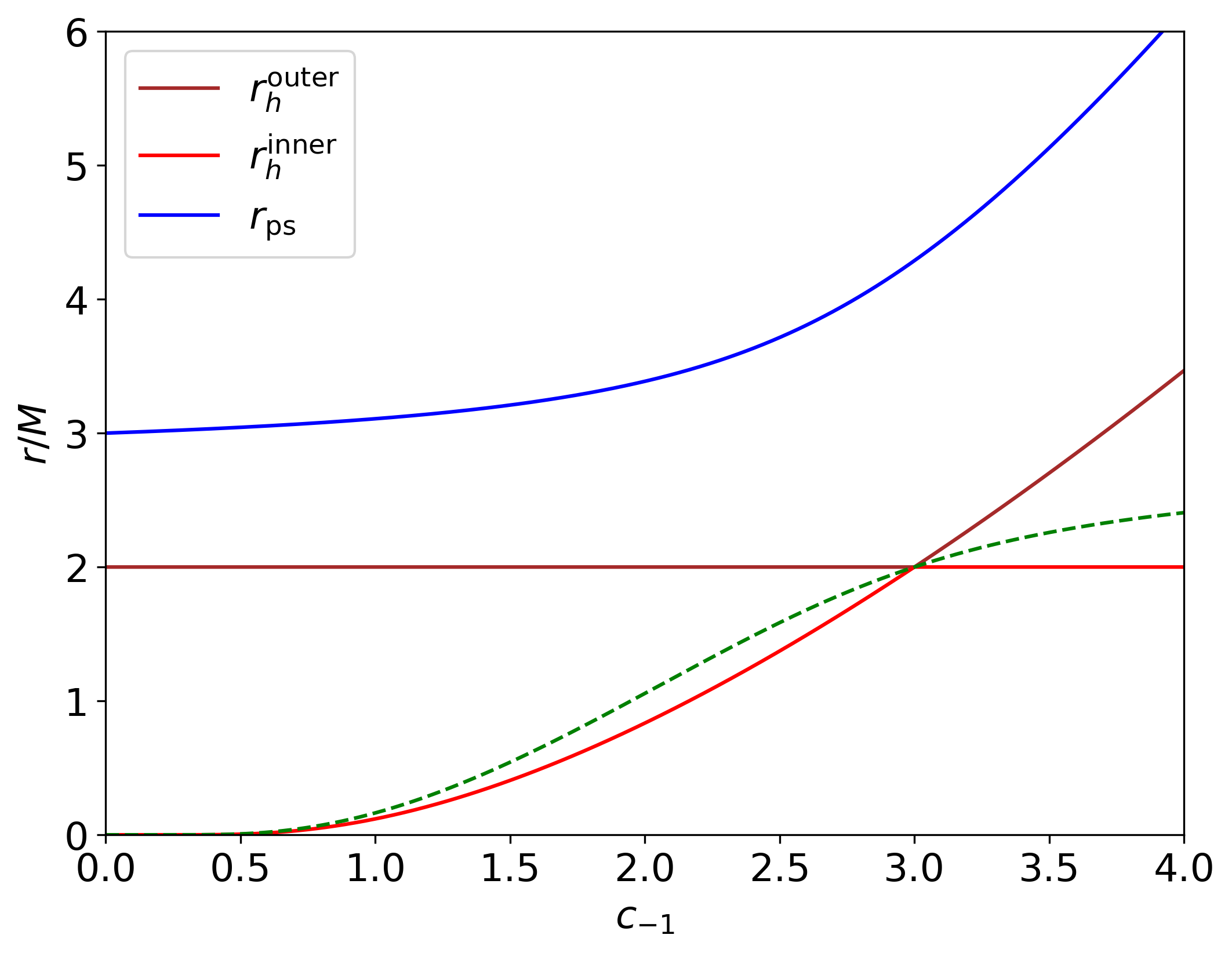}  
    \caption{Outer and inner horizon radii, $r_h^{\rm outer}$ and $r_h^{\rm inner}$,  and the photon sphere radius $r_{\rm ps}$ as functions of the parameter $c_{-1}$  in the scenario with $r_d = \RS$. The green dashed line corresponds to the second root of Eq.~\eqref{eq:ps1}, which lies between the inner and the outer horizons. This  solution does not identify  a physical photon sphere, as such a surface must be timelike.}
    \label{fig:photon_sphere}
\end{figure}

The radii of  circular orbits traced by photons are obtained by solving  $\partial V_{\rm eff}^{\rm null}/\partial r =0$, where the \qm{null} effective potential $V_{\rm eff}^{\rm null}$ can be read off from Eq.  \eqref{eq:veff} with $\alpha =0$. This relation yields
\begin{align}\label{eq:ps1}
    6r+c_{-1}r_d-9\RS-3c_{-1}r_d\ln(r/r_d)=0\,,
\end{align}
whose solutions is
\begin{align}\label{eq:sol_ps}
  r_{\rm ps}=-\frac{c_{-1}r_d}{2}W_{k}\left\{-\frac{2}{c_{-1}}{\rm e}^{\left[1/3-3\RS/(c_{-1}r_d)\right]}\right\}\,,
\end{align}
the argument of  $W$ defining
\begin{align}
x(c_{-1}) = -\frac{2}{c_{-1}}{\rm e}^{\left[1/3-3\RS/(c_{-1}r_d)\right]}.
\end{align}
The function $x(c_{-1})$ is always negative (since $c_{-1}$ is positive),  and satisfies $x(c_{-1})\to 0^-$ as $c_{-1}\to 0^{+}$ or $c_{-1}\to\infty$. The minimum occurs at $c_{-1}=3\RS/r_d$, where
\begin{align}
x_{\rm min} = -\frac{2 r_d}{3 \RS \, {\rm e}^{2/3}}, 
\end{align}
implying that the range is 
\begin{align}
x(c_{-1}) \in \left[x_{\rm min},\, 0\right).
\end{align}
Therefore, following a similar analysis as in Sec. \ref{Sec:event-horizon}, there may be one, two, or no solutions of Eq.~\eqref{eq:ps1}, depending on the value of $c_{-1}$ and the ratio $r_d/\RS$. Typically, the maximum of the effective potential corresponds to the (unstable) photon sphere, while the minimum defines what is known as the stable photon sphere,   sometimes also referred to as the anti-photon sphere~\cite{Olmo:2023lil}. To avoid confusion, throughout the paper we  use the term ``photon sphere'' simply to refer to a real solution of Eq.~\eqref{eq:ps1}. However, in configurations supporting horizon(s), one should recall that only the root lying outside the (outer) event horizon is physically relevant. This follows from the fact that photon spheres are timelike hypersurfaces, while inside the (outer) horizon they    become spacelike (equivalently, inside the outer horizon the coordinates $r$ and $t$ exchange their role). By taking $r_d = \RS$, Fig.~\ref{fig:photon_sphere} provides a visual summary of the  geometry by showing how the photon sphere radius $r_{\rm ps}$ varies with $c_{-1}$, with the corresponding horizon radii $r_h^{\rm outer}$ and $r_h^{\rm inner}$ (cf. Eq. \eqref{event-hor-solution}) also displayed for comparison. 
\begin{figure}[htbp!]
    \centering
    \includegraphics[width=0.85\textwidth]{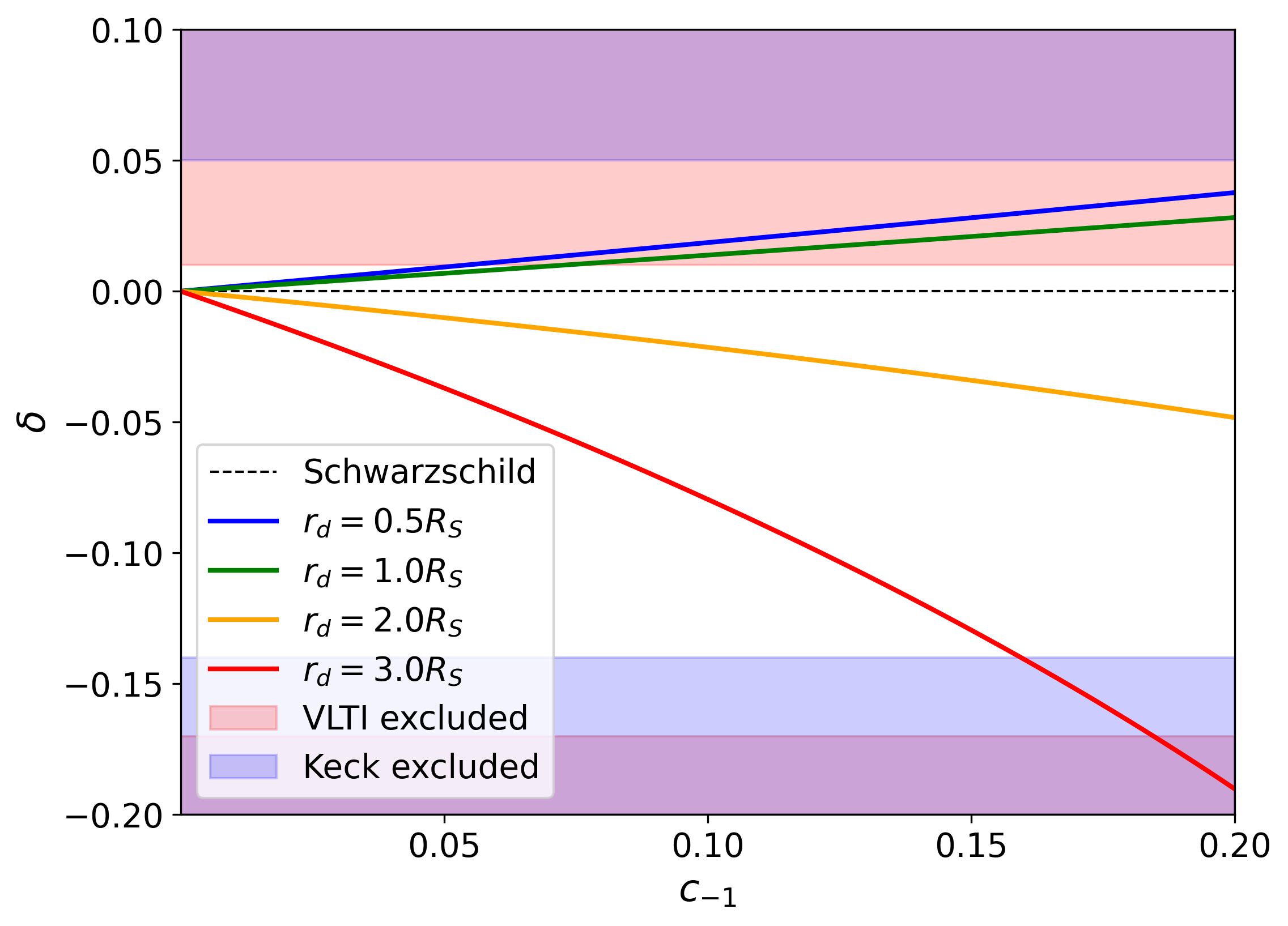}
    \caption{Fractional shadow diameter deviation from the Schwarzschild solution as a function of $c_{-1}$ for different values of the characteristic radius $r_d$. The white region is consistent with the EHT observations of Sgr~A$^\ast$. The geometry \eqref{eq:metric_log} is allowed by empirical constraints for certain ranges of $c_{-1}$.}
    \label{fig:delta_EHT}
\end{figure}

The apparent radius of the photon sphere, typically used to characterize the shadow of a black hole~\cite{Narayan:2019imo,Perlick:2021aok,Afrin2022,Vagnozzi2022c,Vertogradov2024,Wang:2025czc,Wu:2025ihj}, is given by the critical impact parameter~\cite{Vagnozzi:2022moj,Wang:2023rjl,Gao:2023mjb}\footnote{In non-asymptotically flat spacetimes, the critical impact parameter alone is insufficient to fully characterize the observable angular structure~\cite{li2026}; instead, one should compute the apparent angle for an observation at a finite radius. Although our metric does not possess a finite ADM mass, it still approaches the Minkowski form at spatial infinity. Hence, it is valid to use the critical impact parameter alone to study the shadow.}:
\begin{align}
    b_c= \frac{r_{\rm ps}}{\sqrt{B(r_{\rm ps})}}\,,
\end{align}
where, as usual, $b:= L/E$. Defining the fractional deviation parameter as~\cite{EventHorizonTelescope:2022xqj}
\begin{align}\label{eq:delta_b}
   \delta := \frac{b_c}{b_c^{\rm S}}-1\,, 
\end{align}
with $b_c^{\rm S}=3 \sqrt{3} M$ serving as  the Schwarzschild benchmark \cite{Wald}, one can then constrain the factor $c_{-1}$ by comparing the predicted values of $\delta$ with the EHT observations of Sgr~A$^\ast$. The result is presented in Fig.~\ref{fig:delta_EHT}, where the shaded regions are ruled out by the Keck (blue, $\delta = -0.04^{+0.09}_{-0.10}$) and VLTI (red, $\delta = -0.08 ^ {+0.09}_{-0.09}$) priors~\cite{EventHorizonTelescope:2022xqj}.

Remarkably, the logarithmically corrected spacetime \eqref{eq:metric_log} is not excluded by current data and hence can potentially represent a physically viable model. 

Another relevant scale is  defined by the ISCO radius, which plays an important role in the accretion dynamics of  compact objects (see e.g. Refs. \cite{Salahshoor2018,Mustafa2023}). Following a similar procedure as before,  and  imposing that the first and second  derivatives of the effective potential  \eqref{eq:veff} with $\alpha =1$ vanish, we find that $r_{\rm ISCO}$ satisfies 
\begin{align}
&-3 c_{-1}^{\,2} r_d^{\,2}\,
\ln^{2}\!\left(r_{\rm ISCO}/r_d\right)
-2 c_{-1}^{\,2} r_d^{\,2}  
+ c_{-1} r_d \ln\!\left(r_{\rm ISCO}/r_d\right)
\left(4 c_{-1} r_d + 3 r_{\rm ISCO} - 18 \RS \right) \nonumber\\
&\quad
+ 12 c_{-1} r_d \RS
+ 9 \RS \left(r_{\rm ISCO} - 3 \RS\right)
= 0 \,,
\end{align}
which, to first order in  $c_{-1}$, yields
\begin{equation}
r_{\rm ISCO}
=3 \RS+c_{-1}\,r_d
\left[\ln\left(\frac{3 \RS}{r_d}\right)
-\frac{4}{3}\right] +\OO\left(c_{-1}^2\right)\,.    \label{r-ISCO}
\end{equation}

\subsection{Classification of horizon-bearing and horizonless configurations }\label{Sec:BH-mimickers}

Since the log-metric \eqref{eq:metric_log} satisfies all standard energy conditions and is not ruled by EHT data, it is natural to ask whether it can describe the exterior domain of a horizonless compact object which, in some regime,  acts as a black hole mimicker,  or whether it can instead be associated with a new black hole model. In this latter situation, following the terminology adopted in Ref. \cite{Chilingarian2018}, we may refer to it as a   \emph{\qm{bona fide}} black hole,  characterized by an (outer) horizon and a single (physical) unstable photon sphere. 

By contrast, a wide class of horizonless ultra-compact black hole mimickers typically display two light rings: an outer unstable photon sphere  and an inner stable one (the anti-photon sphere) \cite{Cunha:2017qtt,Cardoso-Pani2019,Olmo2021,Olmo:2023lil}. This intriguing setup  naturally leads to a branch structure, with distinct observational consequences for lensing and shadow formation \cite{Olmo:2023lil}, as well as long-lived photon trapping and enhanced time-delay effects, which may be relevant for echo-like features in both electromagnetic and gravitational signals \cite{Guo:2022ghl}.

Introducing the dimensionless parameter
\begin{equation}
X := \frac{3 \RS}{c_{-1} r_d} > 0 ,
\end{equation}
the arguments of the Lambert $W$ functions in Eqs. \eqref{event-hor-solution} and \eqref{eq:sol_ps}   can be recast, respectively, as
\begin{align}
x_h &=
- \frac{3}{c_{-1}}
\ee^{-X},\\
x_{\rm ps} &=
- \frac{2}{c_{-1}}
\ee^{\left(\frac{1}{3}-X\right)}.
\end{align}
Recalling that  $W(x)$ attains real values only for $x\in[-1/\ee,0)$,  the existence of a real photon sphere radius entails that
\begin{equation}
\frac{2}{c_{-1}}
\ee^{\left(\frac{1}{3}-X\right)}
\le \frac{1}{\ee},
\label{cond_ps}
\end{equation}
while the absence of an event horizon is ensured by
\begin{equation}
\frac{3}{c_{-1}}
\ee^{-X}
> \frac{1}{\ee}.
\label{cond_h}
\end{equation}
Combining inequalities \eqref{cond_ps} and \eqref{cond_h}, we obtain that the allowed range of $r_d$ for a fixed  $\RS$ is
\begin{equation}
\frac{3 \RS}
{c_{-1}\!\left[\ln\!\left(\frac{3}{c_{-1}}\right)+1\right]}
\;<\;
r_d
\;\le\;
\frac{3 \RS}
{c_{-1}\!\left[\ln\!\left(\frac{2}{c_{-1}}\right)+\frac{4}{3}\right]} \,.
\label{rd_range}
\end{equation}
Remarkably, within this  interval, the spacetime exhibits one or two photon spheres  while remaining horizonless.

\begin{figure}[htbp!]
    \centering
    \includegraphics[width=0.85\textwidth]{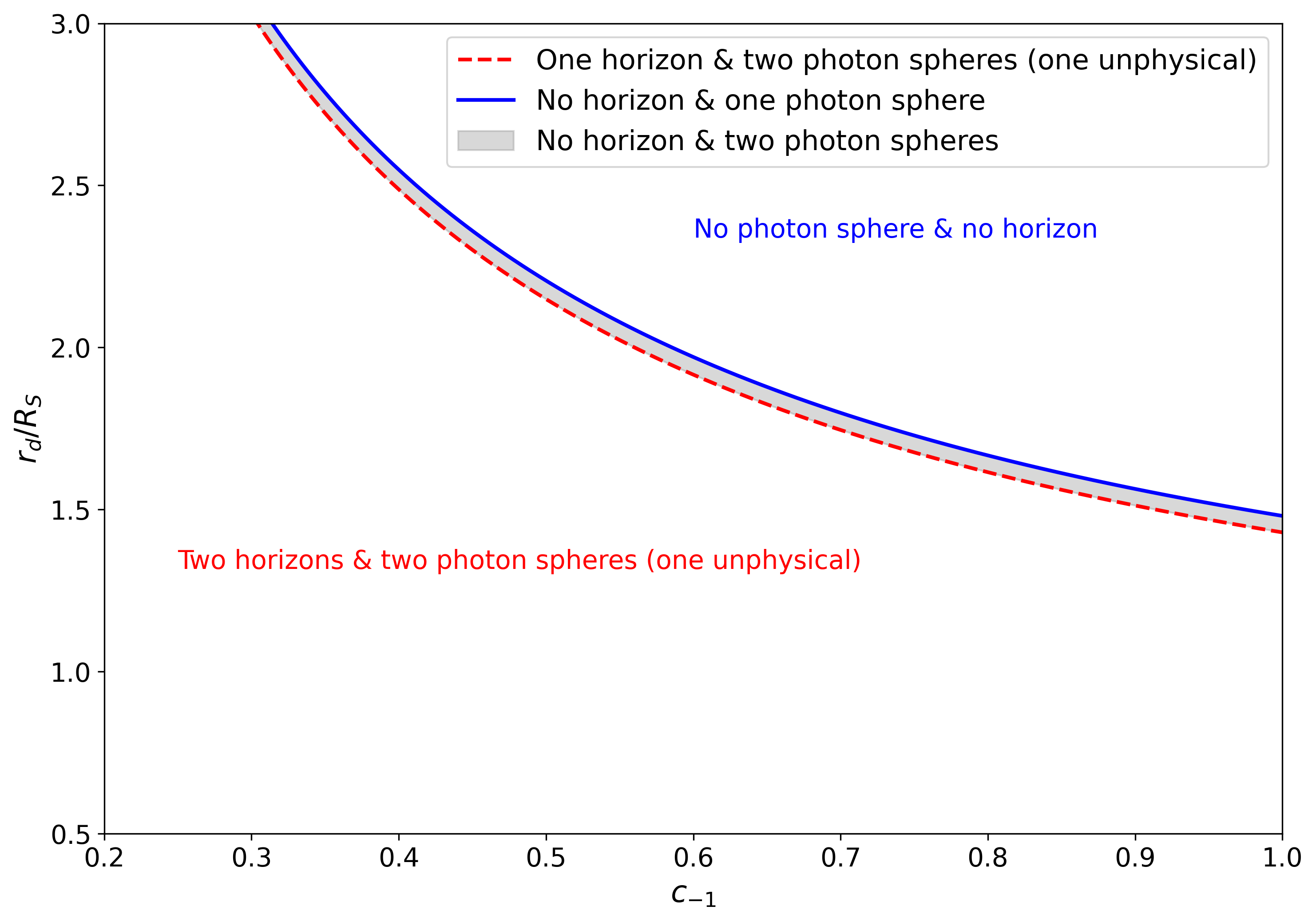}
    \caption{Phase diagram in the $(c_{-1}, r_d/\RS)$ parameter space for the logarithmically corrected geometry \eqref{eq:metric_log}. The red dashed line separates  configurations with two horizons (below) and no horizon (above), while the blue solid line marks the boundary between solutions with two photon spheres (below) and no photon sphere (above). The gray shaded region corresponds to horizonless objects with two photon spheres, which have the key geometric features required to serve as black hole mimickers.}
    \label{fig:phase diagram}
    \end{figure}

The phase diagram in the $(c_{-1}, r_d/\RS)$ parameter space is presented in Fig.~\ref{fig:phase diagram}, while a classification of the potential configurations realized is presented in Table~\ref{tab:configurations-summary}.
    \begin{table*}[h!]
\centering
\renewcommand{\arraystretch}{2} 
\begin{tabular}{|c|c|}
\hline
\textbf{Configuration}  & \textbf{Model} \\ \hline
One horizon, two photon spheres    &      \emph{\qm{Bona fide}} black hole \\ \hline   
Two horizons, two photon spheres    &  \emph{\qm{Bona fide}} black hole   \\  \hline    
No horizon, two photon spheres    &    Black hole mimicker   \\  \hline  
No horizon, one photon sphere    &     \qm{Non-exotic compact object} \\  \hline   
No horizon, no photon sphere    &    \qm{Non-exotic compact object} \\  \hline     
\end{tabular}
\caption{Classification of the logarithmically corrected solution \eqref{eq:metric_log} according to the horizon structure and photon-sphere properties, which are considered in Fig. \ref{fig:phase diagram}. Models featuring one or two horizons and two photon spheres (only one of which is physical) exhibit the geometric properties associated with \emph{\qm{bona fide}} black holes. Horizonless configurations possessing both an unstable  and a stable photon sphere  can be qualified as black hole mimickers, while horizonless solutions with one or no photon sphere can act as \qm{non-exotic compact objects}, as they do not involve  violations of classical energy conditions. }
\label{tab:configurations-summary}
\end{table*}

Horizon-bearing structures possess the key properties  to be  qualified as  \emph{\qm{bona fide}} black holes. Although they always exhibit two extrema of the \qm{null} effective potential   $ V_{\rm eff}^{\rm null}$,  only one  corresponds to a genuine photon sphere, since  the other  lies inside the horizon and is thus unphysical. Moreover, in geometries with two horizons the outer one is  located, in many cases,  at   $r=\RS$ or in its vicinity, the latter scenario arising for sufficiently small values of $c_{-1}$ and arbitrary $r_d$ (cf. Eq.  \eqref{event-hor-solution}). This black-hole-like pattern is  compatible with the occurrence of the $r = 0$ singularity studied in Sec. \ref{Sec:r-0}, so  the matching to a regular interior could be avoided.\footnote{In Ref.~\cite{wang&Battista2026regular}, we will present a detailed discussion on regular geometries, some of which admit regular metrics with logarithmic corrections.}

On the other hand, horizonless models can be  classified  as members of a  new class of compact bodies,  which we  can dub \qm{non-exotic compact objects} to stress that they  respect all classical energy criteria and hence differ from many familiar exotic compact-objects proposals 
(e.g., traversable wormholes or gravastars) that often require energy-condition  violations or rely on quantum effects near the surface \cite{Cardoso-Pani2019}. Horizonless setups supporting two light rings are  characterized by an unstable outer photon sphere and a stable inner   anti-photon sphere. These frameworks can thus potentially act as  black hole mimickers, provided that they reproduce the exterior phenomenology of a black hole  at the level of observables. This construction requires,  as  pointed out before,  interpreting the log-corrected metric  as an effective exterior geometry  joined to a regular interior that  removes the (naked) $r=0$ singularity, which can nonetheless  be  reached only by radial null geodesics (see Sec. \ref{Sec:r-0}).  This viewpoint also facilitates drawing a parallel with exotic compact objects,  many of which are explicitly designed  to avoid a singular central region by means of similar matching prescriptions.

In order to further explore  the possibility of associating the  solution \eqref{eq:metric_log} with the description of compact objects, we recall that  the  compactness of a gravitating body with mass $\mathcal{M}$ and  finite characteristic radius $R$ is conventionally defined as $ \mathcal{C} := \mathcal{M}/R$ \cite{Rosswog2014}. Configurations with   $ \mathcal{C}  \sim \OO(1)$ are typically regarded as highly compact \cite{DeLaurentis2025}  (e.g., for neutron stars one generally finds $\mathcal{C} \sim 0.2$, which is orders of magnitude higher than  the solar-system benchmark value  $\mathcal{C} \sim 10^{-6}$  \cite{Rosswog2014}), although in general relativity  the Buchdahl theorem imposes the upper bound $\mathcal{C} \leq 4/9$   \cite{Cardoso-Pani2019}. Based on the taxonomy outlined  in Ref. \cite{Cardoso-Pani2019}, our log-geometry  displays the essential characteristics to pertain to  compact or even ultra-compact objects, as it admits an ISCO that can be located at $r_{\rm ISCO} \sim 3 \RS$ (cf. Eq. \eqref{r-ISCO}) and a photon sphere whose radius can be read off from Eq. \eqref{eq:sol_ps} (see Fig. \ref{fig:photon_sphere} for the case $r_d=\RS$ in horizon-bearing setups). 

An additional indicator of high compactness fulfilled by the log-metric is the existence of  infinite-redshift surfaces or  \emph{nearly} infinite-redshift surfaces  (see e.g. Refs. \cite{Ovalle2023,Casadio2024} and references therein). To illustrate this point, we focus on the special parameter choice for which the metric function $B(r)$
admits a degenerate zero at $r=r_h$, namely \begin{equation}\label{eq:doubleroot}
B(r_h)=0, \qquad B'(r_h)=0 .
\end{equation}
\begin{figure}[htbp!]
    \centering
    \includegraphics[width=0.7\textwidth]{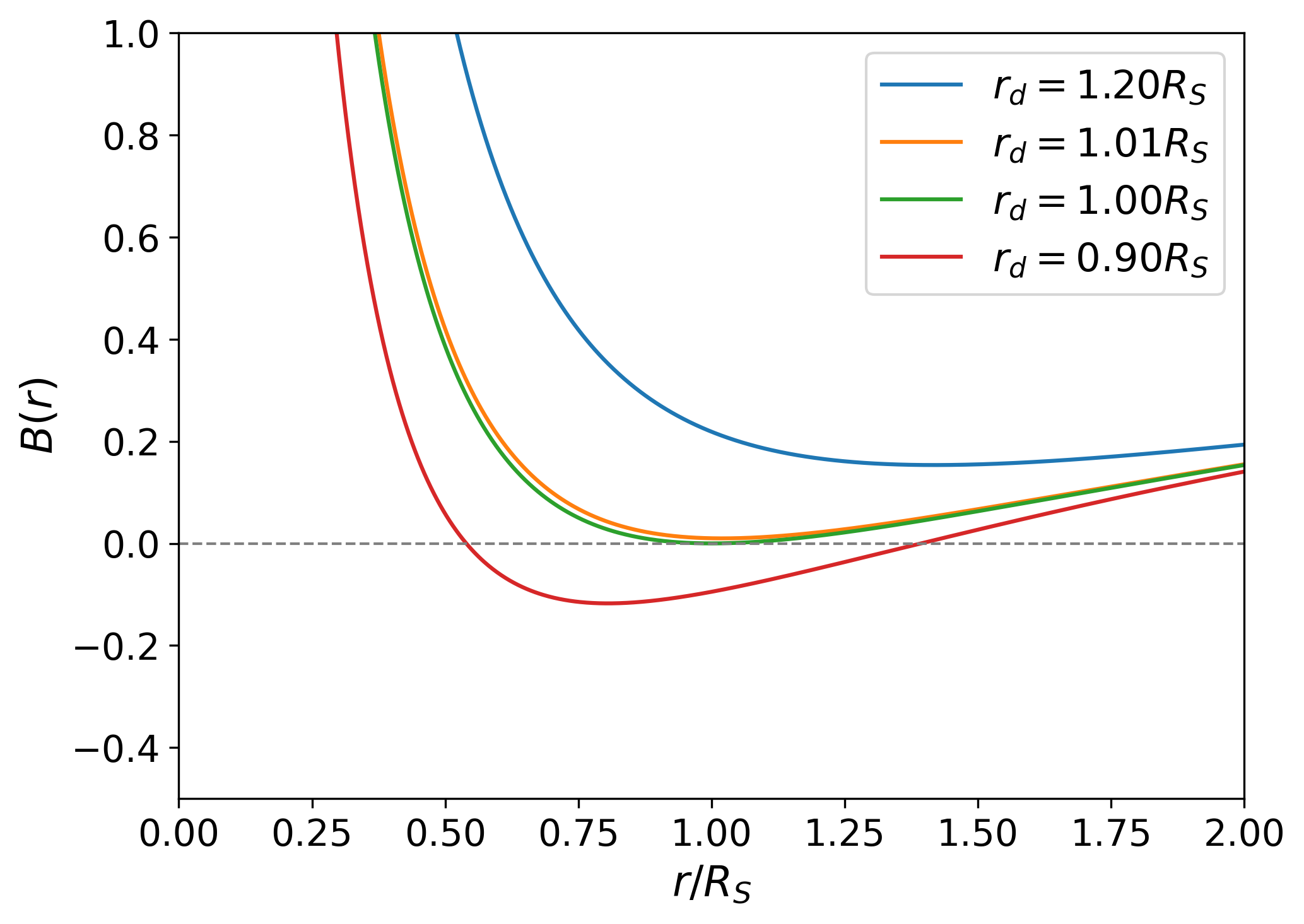}
    \caption{The metric function $B(r)$ for the logarithmically corrected spacetime with fixed $c_{-1}=3$ and selected representative values of the scale parameter $r_d$. The threshold case $r_d=\RS$ (green curve) corresponds to a degenerate root at $r=r_h$ (cf. Eq.  \eqref{eq:doubleroot}), signaling the emergence of an infinite-redshift surface. Small deviations from this limiting scenario remove the degeneracy and  either produce a \emph{\qm{bona fide}} black hole with two horizons (red curve), or lead to a nearly infinite-redshift surface (more pronounced for the orange curve and slightly less  for the blue curve), thereby maintaining the high compactness of the configuration. } 
    \label{fig:UCO}
    \end{figure}
Solving these conditions yields 
\begin{equation}
c_{-1}=\frac{3r_h}{r_d}, 
\qquad 
r_h\!\left[1-\ln\!\left(\frac{r_h}{r_d}\right)\right]=\RS\,,
\end{equation}
indicating that $r=r_h$ is a double root of $B(r)$. Consequently, the redshift factor $z$ for signals emitted by static sources approaching $r_h$ diverges as
\begin{equation}
z \sim B(r)^{-1/2} \;\longrightarrow\; \infty,
\qquad r\to r_h^+ .
\end{equation}
Thus,  $r=r_h$  corresponds to an infinite-redshift surface (recall that the event horizon of a static,  spherically symmetric spacetime is automatically an infinite-redshift surface). Conversely, scenarios where $B(r)$ never vanishes but can be arbitrarily close to zero (as displayed by the orange curve and,  to a lesser extent, by the blue curve in Fig. \ref{fig:UCO}) give rise to horizonless  patterns with a nearly infinite-redshift surface  at $r=r_z$. In these cases, the exact relations  \eqref{eq:doubleroot} are no longer strictly satisfied, but one finds
\begin{equation}
B(r_z)\to 0^+, \qquad B'(r_z)=0\,,
\end{equation}
which describe signals experiencing extremely large, albeit finite, gravitational redshift at $r=r_z$. 
The behavior of  $B(r)$ for $c_{-1}=3$ is  shown in Fig. \ref{fig:UCO}, which illustrates that the geometry allows for both infinite- and nearly infinite-redshift surfaces.

Some final remarks are now in order. Another significant facet  supporting the interpretation of Eq. \eqref{eq:metric_log} as a suitable effective exterior geometry is  that this metric is unlikely to originate from standard gravitational-collapse processes, for at least two reasons: (i) timelike geodesics are prevented from reaching $r=0$; (ii) this curvature singularity persists also in the absence of an event horizon (see Sec. \ref{Sec:r-0}).  Moreover,  at this stage, the solution should be regarded as a \emph{geometric} rather than  astrophysical candidate for a new family of compact bodies, possibly including black hole mimickers. Our analysis is indeed meant to be  preliminary, since key quantities  such as  the closeness parameter, as well as  important aspects like  formation  and stability mechanisms,  have not yet been  assessed \cite{Delgaty1998,Visser2003b,Cardoso-Pani2019,Herdeiro2022,Bezares2024,Carballo-Rubio:2021bpr}. Nonetheless, our examination shows that Eq. \eqref{eq:metric_log}  contains all the essential ingredients for modeling such compact configurations,  providing further evidence  that  the algorithm devised in Sec. \ref{Sec:NEC-constraints-AB-1},    based on Eqs. \eqref{y-condition} and \eqref{eq:def_y}, has a solid physical grounding.

\subsection{Constraints from solar-system-like environments}\label{Sec:solar-system-tets}

In the spirit of treating the log-corrected spacetime  as an effective exterior geometry, we now investigate bounds on $c_{-1}$ stemming from solar-system-like environments,  which probe the metric in the weak-field regime and hence are insensitive to its behavior  at small radii or to the details of a possible interior completion. From this perspective,  all  constraints derived here should not be interpreted as fundamental theoretical restrictions on the model, nor do they supersede the less stringent limitations  provided by  EHT observations discussed in Sec. \ref{Sec:photon-sphere}. 

We begin by analyzing the gravitational redshift in Sec. \ref{Sec: grav-redshift}. Then, the bending of light and the orbit precession are explored in Secs. \ref{Sec:bending-light} and \ref{Sec:Precession-Mercury}, respectively.

In this section,  we set the length scale $r_d$ to be equal to $\RS$.

\subsubsection{Gravitational redshift} \label{Sec: grav-redshift}

The gravitational redshift relates the emitted and received frequencies, $\omega_1$ and $\omega_2$, for two static observers located at radii $r_1$ and $r_2$, via \cite{Wald}
\begin{align}
    \frac{\omega_1}{\omega_2}=\frac{\sqrt{B(r_2)}}{\sqrt{B(r_1)}}\,.
\end{align}
In the limit $\RS/r\ll 1$,  we find 
\begin{align}
    \frac{1}{B(r)}=&1+\frac{[3+c_{-1}\ln(r/\RS)]\RS}{3r}\notag \\
    &+\frac{[9+6c_{-1}\ln(r/\RS)+2c_{-1}^2\ln(r/\RS)]\RS^2}{9r^2}+...\,,
\end{align}
where we have regarded $\ln(r/\RS)$ as a zeroth-order term in the expansion in $\RS/r$. At leading order, the fractional frequency shift $\delta_\omega:=\left(\omega_2-\omega_1\right)/\omega_1$  becomes
\begin{align}\label{eq:def_redshift}
   \delta_{\omega} \approx\Phi(r_1)-\Phi(r_2)\,,
\end{align}
where  $\Phi$ is the effective Newtonian gravitational potential in the weak-field limit:
\begin{align}
    \Phi=\frac{B(r)-1}{2}= -\frac{\RS}{2r}\left[1+\frac{c_{-1}\ln(r/\RS)}{3}\right]\,.
\end{align}
Owing to  Eq.~\eqref{eq:def_redshift}, the following relation for $c_{-1}$ can be derived:
\begin{align}
    c_{-1} = 
\frac{
3\bigl[ 2\, r_1 r_2 \,\delta_{\omega}+ \RS (r_2 - r_1) \bigr]
}{
\RS \bigl[ r_1 \ln (r_2/\RS) - r_2 \ln (r_1/\RS) \bigr]
}.
\end{align}

The gravitational redshift has been experimentally confirmed with an accuracy of approximately $2\times 10^{-4}$ by the  space-based  experiment Gravity Probe A~\cite{Will:2014kxa}, assuming the Schwarzschild solution.  Taking $r_1$ to be the Earth radius  $R_{\oplus}\approx 6371$ km,  $\RS\approx 8.87\times 10^{-6}$ km the corresponding Schwarzschild radius, and $r_2\approx 16371$ km the radial coordinate of the receiver located at an altitude   of about $10000$ km above the Earth surface (reflecting the standard setup used in Gravity Probe A tests~\cite{Will:2014kxa}),  we arrive at the solar system constraint
\begin{align}
    \left. c_{-1}\right \vert_{\rm solar\, system} ^{\rm redshift}\lesssim  3.03\times 10^{-5}\,.
\end{align}

\subsubsection{Bending of light}\label{Sec:bending-light}

Starting from the geodesic equations \eqref{eq:geodesic_log} and  \eqref{eq:veff} with $\alpha=0$,  together with the definition \eqref{eq:defL} of angular momentum, one can easily show that the spatial  orbit of a photon in the equatorial plane of the spacetime \eqref{eq:metric_log} is governed by 
\begin{align}
\frac{\dd \phi}{\dd r}=\pm\frac{1}{r^2}\left[\frac{1}{b^2}-\frac{B(r)}{r^2} \right]^{-1/2}\,,
\label{bending-eq-1}
\end{align}
where we recall $b:= L/E$ and the upper (resp. lower) sign refers to clockwise (resp. counterclockwise) motion. 

The deflection of the light trajectory from a straight line is defined  as $\delta \phi:=\Delta \phi-\pi$, where  $\Delta \phi \equiv  |\phi_{+\infty}-\phi_{-\infty}|$  is the total change of the angular coordinate $\phi$,  with $\phi_{-\infty}$ and  $\phi_{+\infty}$  the asymptotic values  along the incoming and outgoing photon paths, respectively. Since at the distance of closest approach $r_0$ the quantity  $\dd r/\dd \phi$ vanishes, Eq. \eqref{bending-eq-1} gives
\begin{align}\label{eq:b_r0}
    b^2=\frac{r_0^2}{B(r_0)}\,,
\end{align}
and hence, by expanding the integrand of the relation  $\Delta\phi=2\int\limits_{r_0}^{\infty}\dd r \, r^{-2}\left[B(r_0)/r_0^2-B(r)/r^2\right]^{-1/2}$ in the small parameter $\RS/r$,  we obtain, upon neglecting higher-order contributions,
\begin{align}\label{eq:Delta phi-2}
\Delta\phi=&2\int\limits_{r_0}^{\infty}\frac{\dd r}{r\left[{r^2}/{r_0^2}-1\right]^{1/2}}\Bigg[1+\frac{(r^2+rr_0+r_0^2)\RS}{2r_0(r+r_0)r}\notag \\
&+\frac{ c_{-1}\left[ r^3 \ln \left({r_0}/{\RS}\right)- r_0^3 \ln \left({r}/{\RS}\right)\right]\RS}{6r_0 \left(r^2-r_0^2\right)r}+...\Bigg]\,.
\end{align}
Performing the integral yields 
\begin{align}
     \delta\phi= \frac{2\RS}{r_0}\left[1+\frac{c_{-1}}{3}\ln\frac{r_0}{2\RS} \right]+...\,,
\end{align}
which translates into the general expression for the coefficient $c_{-1}$:
\begin{align}\label{c-1-general-bend}
c_{-1}= \frac{3}{\ln (r_0/2\RS)}\left(\frac{r_0 \delta \phi }{2\RS}-1 \right).
\end{align}

In the case of a light ray grazing  the solar surface, we can use $\RS\approx 2.95$ km and  $r_0=R_{\odot}\approx6.96\times10^5$ km. Since  very-long-baseline  interferometry measurements   agree with the general-relativistic prediction  ${2\RS}/R_{\odot}$ to an accuracy of 0.01 $\%$~\cite{Will:2014kxa}, we can conclude from Eq.  \eqref{c-1-general-bend} that 
\begin{align}
   \left. c_{-1}\right \vert_{\rm solar\, system} ^{\rm light \, bending}\lesssim 2.57\times 10^{-5}\,.
\end{align}

\subsubsection{Orbit precession}\label{Sec:Precession-Mercury}

In the previous section, we have dealt with the unbounded trajectories of photons. We now turn to the bound orbits  of  massive particles around a central object. Following standard arguments \cite{Weinberg1972}, one finds that this motion is ruled by
\begin{align}\label{eq:mu_phi}
    \left(\frac{\dd \mu}{\dd \phi}\right)^2+B\left(\mu^2+\frac{4L^2}{\RS ^2}\right)=\frac{4E^2 L^2}{\RS ^2}\,,
\end{align}
where we have introduced the  dimensionless parameter~\cite{Carroll2004}
\begin{align}\label{eq:def_mu}
    \mu :=\frac{2L^2}{ \RS r}\,.
\end{align}
Differentiating Eq. \eqref{eq:mu_phi} with respect to $\phi$   in the limit  $\RS/r\sim\mu\RS^2/L^2 \ll 1$, we obtain,  for our explicit form of $B$ given in  Eq. \eqref{eq:metric_log}, 
\begin{align}\label{eq:mu_full}
    \frac{\dd^2\mu}{\dd \phi^2} +\mu-1=h\left(\mu,\RS,c_{-1}\right)\,,
\end{align}
where the function
\begin{align}\label{eq:defh}
    h\left(\mu,\RS,c_{-1}\right)\equiv \frac{3\RS^2\mu^2}{4L^2}-\frac{c_{-1} }{3}\ln\left(\frac{\mu \RS ^2}{2L^2}\right)+...\,,
\end{align}
collects the leading relativistic and logarithmic corrections.
 In  Newtonian gravity, where $h$ vanishes, we get a solution describing a perfect ellipse with eccentricity $e$:
\begin{align}\label{eq:mu0_sol}
    \mu_0=1+e\cos \phi\,.
\end{align}
We can thus assume the perturbative expansion of $\mu$ 
\begin{align}
    \mu=\mu_0+\mu_1\,,
\end{align}
where $\mu_1$ represents a small deviation from the Newtonian trajectory $\mu_0$. Substituting this \emph{ansatz} into Eq.~\eqref{eq:mu_full} and retaining only first-order corrections in $\mu_1$, we obtain
\begin{align}
    \frac{\dd^2\mu_1}{\dd \phi^2} +\mu_1=h(\mu_0,\RS,c_{-1})\,,
\end{align}
which admits an analytical solution, although the full expression is rather lengthy. For our purposes, the most relevant term is the non-constant and non-periodic (dominant secular) contribution
\begin{align}
    \mu_1=\left(\frac{3e\RS^2}{4L^2}-\frac{c_{-1}}{3e}\right)\phi\sin \phi+...\,,
\end{align}
governing the cumulative precession. Therefore, the orbit is described by
\begin{align}\label{eq:mu_persol}
    \mu=1+e\cos \phi+\left(\frac{3\RS^2}{4L^2}-\frac{c_{-1}}{3e^2}\right)e\phi\sin \phi\,.
\end{align}
If we introduce the small parameter 
\begin{align}
    \gamma :=\frac{3\RS^2}{4L^2}-\frac{c_{-1}}{3e^2}\,,
\end{align}
then Eq.~\eqref{eq:mu_persol} can be formulated as an ellipse with a period different from $2\pi$:
\begin{align}
    \mu=1+e\cos\left[(1-\gamma)\phi\right]\, ,
\end{align}
and thus the perihelion precession per orbit is  given by $\Delta \phi_{\rm prec}=2\pi \gamma$. Bearing in mind the standard polar equation of the ellipse with semi-major axis $a$
\begin{align}
    r(\phi)=\frac{(1-e^2)a}{1+e\cos\phi}\,,
\end{align}
jointly with the definition of $\mu$ from Eq.~\eqref{eq:def_mu} and the zero-order solution \eqref{eq:mu0_sol}, we arrive at the approximate formula for the angular momentum
\begin{align}
    L^2\approx a(1-e^2)\RS/2\,.
\end{align}
The perihelion advance per orbit  then reads as
\begin{align}
    \Delta \phi_{\rm prec}= \frac{3\pi\RS}{(1-e^2)a}\Lambda\,,
\end{align}
where the dimensionless parameter $\Lambda$ encodes departures from standard Schwarzschild metric 
\begin{align}
  \Lambda\equiv 1-\frac{2c_{-1}a(1-e^2)}{9e^2 \RS}\,,
\end{align}
which solving for $c_{-1}$ leads to 
\begin{align}\label{eq:c_perihelion}
    c_{-1}=\frac{(1-\Lambda)9e^2\RS}{2a(1-e^2)}\,.
\end{align}
This formula facilitates direct comparison with established post-Newtonian constraints. Using the results from Refs.~\cite{will2018theory,Clifton:2005aj,shapiro1976verification}
\begin{align}
    \Lambda=1.003\pm0.005\,,
\end{align}
together with relevant orbital elements of Mercury, $a\simeq5.79\times 10^7$ km and $e=0.2056$ \cite{Weinberg1972}, we infer the limit 
\begin{align}
    \left. c_{-1}\right \vert_{\rm solar\, system}^{\rm precession}\lesssim 2.02\times 10^{-11}\,.\label{precession-bound-1}
\end{align}

Apart from  the  above weak-field solar-system bound, a complementary mid-field constraint can be obtained from the relativistic precession of the S2 star orbiting  Sgr~A$^\ast$, for which the GRAVITY Collaboration  has reported a perihelion advance of approximately $12'$ per orbital period~\cite{GRAVITY:2020gka}. Following Ref. \cite{GRAVITY:2020gka}, which neglects Kerr spin effects,  deviations from the Schwarzschild prediction can be quantified via the  dimensionless parameter $\Lambda_{\rm S2}$ (denoted by  $f_{\rm SP}$ in Ref. \cite{GRAVITY:2020gka}), whose established value is
\begin{align}
  \Lambda_{\rm S2} =1.10\pm0.19\,.
\end{align}
Using the S2 orbital parameters $a\simeq125~\RS$ and $e\simeq 0.885$~\cite{GRAVITY:2020gka}, Eq.~\eqref{eq:c_perihelion} yields the  bound
\begin{align}\label{eq:S2}
    \left.c_{-1}\right\vert_{\rm S2}^{\rm precession}\lesssim 0.0117\,,
\end{align}
which is several orders of magnitude weaker than the solar-system limit \eqref{precession-bound-1}. The milder constraint suggests that the log-corrected geometry~\eqref{eq:metric_log} is  potentially more relevant for describing galactic environments  than  the solar system. Importantly, the bound~\eqref{eq:S2} lies entirely within the parameter region allowed by EHT observations of Sgr~A$^\ast$, as illustrated in Fig.~\ref{fig:delta_EHT}.

\subsection{Junction conditions}\label{Sec:junction-condition}

Although asymptotically flat (cf. Eq. \eqref{eq:asy_flat}), the  Misner-Sharp-Hernandez mass  $\mathscr{M}=\tfrac{r}{2}\left(1-\nabla^\mu r \nabla_\mu r\right)$  associated with the spacetime \eqref{eq:metric_log} grows logarithmically and hence diverges at infinity, a behavior  known to be pathological for isolated systems \cite{Faraoni2021} (in other words, the metric is not  asymptotically Schwarzschild in the standard \qm{isolated system} sense). To cure this issue and give a consistent physical interpretation to the  model, we assume, in line with its effective character, that beyond some large radius $\mathcal{R}>2M_{\infty}$  the geometry is joined to an exterior Schwarzschild region
\begin{align}
    \dd s^2_{\rm S}=-\left(1-\frac{2M_{\infty}}{r}\right) \dd t^2+ \left(1-\frac{2M_{\infty}}{r}\right)^{-1} \dd r^2 + r^2 \dd \Omega^2. 
\end{align}

Let $\Sigma$ be the timelike hypersurface  $r=\mathcal{R}$, where we install the intrinsic coordinates $y^A=\left(\tau,\theta,\phi\right)$ (here,  capital Latin indices $A,B=1,2,3$), with $\tau$ the proper time of  static observers  on $\Sigma$. Then, following the general recipe of the Israel-Lanczos formalism \cite{visser1995lorentzian,Poisson2009},  we find that the first Darmois junction condition, i.e., the induced three-metric $h^{AB}$ having no jump discontinuity across $\Sigma$, requires that  temporal (and thus radial) metric components be the same on both sides of  $\Sigma$, which thus yields the relation (recall that $\RS \equiv 2M$)
\begin{align}
M_\infty= M + \frac{c_{-1}r_d}{6} \ln\left(\mathcal{R}/r_d\right).\label{M-infty} 
\end{align}

Therefore,  $M_\infty$ represents the ADM mass, i.e., the  total physical (\qm{renormalized}) mass measured at infinity by an asymptotic observer after the matching has been performed. On the other hand, $M$ can  be regarded as an effective (\qm{bare}) interior mass  associated with the pure $1/r$ contribution to $B(r)$. This parameter is involved in the description of the gravitational field (it enters, for instance, the equations determining the horizons, photon spheres, and ISCOs) in the exterior region $\mathcal{R}_0<r<\mathcal{R}$, where $\mathcal{R}_0$ denotes the radius of the inner (possibly regular) completion. For instance, in the solar-system-like environments considered in Sec. \ref{Sec:solar-system-tets}, the difference between $M_\infty$ and $M$ can be extremely small as $c_{-1}\ll 1$. 

The second junction condition, i.e., continuity of the extrinsic curvature, demands the equality of the radial derivatives of temporal metric components at $\Sigma$. This  criterion is not fulfilled owing to  the  mismatch
\begin{align}
\left.\left(1-\frac{2M_\infty}{r}\right)'\right\vert_\Sigma-\left.\left(1-\frac{2M}{r}-\frac{c_{-1} r_d\ln (r/r_d)}{3r}\right)'\right\vert_\Sigma=\frac{c_{-1}r_d}{3 \mathcal{R}^2},
\end{align}
where we have used Eq. \eqref{M-infty}. This situation is interpreted physically as the presence of a surface layer (or thin shell) at $\Sigma$ having a distributional stress-energy tensor.

The corresponding surface stress-energy tensor $\mathscr{S}^{AB}$ can be expressed in the perfect-fluid form \cite{Visser2003b,Poisson2009,Berry2020}
\begin{align}
\mathscr{S}^{AB} = \left(\sigma+ \mathscr{P}\right)  \mathscr{U}^A \mathscr{U}^B + \mathscr{P}h^{AB},
\end{align}
where $\sigma$ and $\mathscr{P}$ denote the surface energy density and pressure, respectively,  and $\mathscr{U}^A$ the velocity of a static observer sitting at $\Sigma$. In our setup, it is not complicated to show that \footnote{The dimension of $\mathscr{P}$ is $\sim 1/\mathcal{R}$ (or $L^{-1}$), whereas from Eq.~\eqref{eq:rhoP}, the dimension of $p_2$ is $\sim 1/r^2$ (i.e., $L^{-2}$). This difference stems from the fact that $\mathscr{P}$ is a surface pressure defined on a two-dimensional timelike hypersurface, while $p_2$ is the usual bulk pressure defined in three-dimensional space.} 
\begin{align}
\sigma=& 0,
\\
\mathscr{P}=&\frac{1}{16\pi} \left(1-\frac{2M_\infty}{\mathcal{R}}\right)^{-1/2}\frac{c_{-1}r_d}{3 \mathcal{R}^2}.
\end{align}

$ \mathscr{S}^{AB}$ complies with the NEC, WEC, and SEC, and therefore does not call for exotic matter in the usual sense,  while DEC is spoiled (since $\sigma=0$ and $\mathscr{P}>0$). However, the tangential pressure $\mathscr{P} \sim c_{-1} r_d /\mathcal{R}^{2} $ and thus it can be arbitrarily small as $c_{-1}\ll 1$ and $r_d/\mathcal{R} \ll 1$, which means that the resulting  DEC violation is extremely mild. This situation is  significantly better than in many thin-shell models where DEC and  even weaker conditions fail (e.g., in Refs. \cite{Poisson1995,Sharif2014,Berry2020} a thin layer with a negative surface energy density is considered), although DEC-satisfying constructions exist, like e.g. in Ref. \cite{Brady1991}.

Nevertheless, a different approach can also be pursued (this is related to so-called thick-shell or finite-width layers generalizations; see e.g. Refs. \cite{Cattoen2005,Berry2022,Jampolski2023,DiFilippo2024,Rahaman2026}). While a sharp junction at $r = \mathcal{R}$ 
necessarily produces a surface layer, performing the matching across a narrow region  $r \in [\mathcal{R}, \mathcal{R}+\Delta]$,  with $\Delta \ll \mathcal{R}$, allows the metric  to be smoothly interpolated  between the logarithmic solution and the Schwarzschild exterior  using  an appropriately chosen smooth function (e.g., a quintic polynomial). This construction can ensure continuity of both the metric and its first derivative, and hence can yield a fully regular matching procedure compatible with the classical energy conditions.

Before concluding, a last crucial remark is in order. One may notice that  the log-corrected geometry  \eqref{eq:metric_log} may be recast in the one-parameter form  via the identity 
$-g_{tt}=1-\RS/r-c_{-1} r_d\ln (r/r_d)/(3r)=1-\tilde{R}_{\rm S}/r- \tilde{\lambda}\ln (r/\tilde{\lambda})/r$, with $\tilde{\lambda}:= c_{-1}r_d/3$ and $\tilde{R}_{\rm S} :=\RS+\tilde{\lambda} \ln (c_{-1}/3)$. Incidentally, this reproduces the perfect-fluid dark-matter black hole model considered in Ref.~\cite{Liang:2023jrj} (see Eq. (2.11) therein).  Nevertheless, the physical picture is different in the present context for various reasons,  and Eq. \eqref{eq:metric_log} is more naturally regarded as a two-parameter solution. First, let us recall that $c_{-1}$ is dimensionless and  $r_d$  dimensionful,   while  the reparametrization in terms of $\tilde{\lambda}$ mixes them  into a single dimensionful quantity. This reflects the fact that $c_{-1}$ and $r_d$ arise independently in our framework, where they fulfill distinct physical roles:  the former controls the departure from the vacuum Schwarzschild limit, which is recovered for $c_{-1}=0$, whereas  the latter  merely represents a  length scale. Moreover, as discussed above,   $M = \RS/2$ admits a clear physical interpretation as  an effective  interior mass, related to the ADM mass via Eq. \eqref{M-infty}. On the other hand, $\tilde{M} = \tilde{R}_{\rm S} /2$ does not share this meaning and coincides with  $M$ only for the special case $c_{-1}=3$. If the logarithmic metric is interpreted as an effective description of a dark matter halo, then the adopted form $-g_{tt}=1-\RS/r-c_{-1} r_d\ln (r/r_d)/(3r)$ naturally singles out two distinct pieces: a central point-mass term and a surrounding dark-matter-like halo correction. Conversely, in the expression $1-\tilde{R}_{\rm S}/r- \tilde{\lambda}\ln (r/\tilde{\lambda})/r$ these  contributions become mixed (the mixing only disappears  when $c_{-1}=3$), and their individual physical origin is obscured. However, the solution \eqref{eq:metric_log} is sufficiently general that,  upon choosing the scale $r_d$ as  the halo radius,  our component $B(r)$ reproduces the metric function given by Eq. (3.13) in Ref.~\cite{Liang:2023jrj}.  Within the above matching procedure, this identification is  equivalent to setting $\mathcal{R} = r_d$.

\section{Concluding remarks}\label{Sec:conclusions}

Classical energy conditions are essential  to guarantee that Einstein field equations are sourced by \qm{physically reasonable} matter and that the resulting solutions are physically admissible. 
In this work, we have examined their role in generic static, spherically symmetric spacetimes \eqref{eq:metric_static}.  

In models where $-g_{tt} \neq g_{rr}^{-1}$ (see Sec.~\ref{Sec:A-B-neq-const}), the formation of an event horizon is often accompanied by NEC violations, and  some classical expectations may break down. Such scenarios can lead to the emergence of  effective impenetrable surfaces characterized by $-g_{tt} g_{rr} \equiv AB = 0$, where  the stress-energy tensor typically exhibits a pathological behavior, with the energy density and pressures becoming divergent. Interestingly,  this pattern  is reminiscent of the firewall hypothesis \cite{Almheiri:2012rt}, which proposes that infalling observers may encounter a high-energy surface and be destroyed upon reaching the black hole horizon.

We also notice that although expressions like $A(r)B(r)$ are not scalar objects under coordinate transformations,  the main physical conclusions drawn in this paper are coordinate-independent.  In particular, the energy conditions are defined through covariant contractions of the energy-momentum tensor with causal vectors; the limit $AB\to0$ at $r>0$ corresponds to $\det g_{\mu\nu}\to0$ and hence to a genuine degeneracy of the metric \eqref{eq:metric_static};  the unboundedness of the expansion scalar $\vartheta$ (cf. Eq. \eqref{eq:expansion}) characterizes a coordinate-independent  focusing behavior of a geodesic congruence,  once this is specified;  the divergences of quantities measured in freely falling  frames signal true physical effects, rather than merely coordinate artifacts.

Within the class  of Kerr-Schild  spacetimes \eqref{eq:metric_static-3} (see Sec. \ref{Sec:A-one-over-B}), we have developed a systematic procedure for deriving metrics obeying the NEC, where one simply selects the function $y(r)$ fulfilling relation   \eqref{y-condition} and then solves the ensuing \emph{ordinary} differential equation \eqref{eq:def_y} for the component $B(r)$.

Among the several solutions that can be constructed via this algorithm, we have identified in Sec. \ref{Sec:geometry-log-correction}  a particularly notable example that includes a logarithmic correction to the Schwarzschild model. Similar configurations  have appeared in modified gravity patterns, but our  Eq. \eqref{eq:metric_log} is  distinctive in that it is framed within general relativity and  satisfies \emph{all} standard energy conditions. We have analyzed its parameter space by distinguishing between different classes of compact and ultra-compact objects according to the presence or absence of horizons and photon spheres, leading to the following operational classification (see Fig. \ref{fig:phase diagram} and Table \ref{tab:configurations-summary}): (i) \emph{\qm{bona fide}} black holes, characterized by the existence of at least one  event horizon and a physical photon sphere; (ii) black hole mimickers, i.e., horizonless structures  typically possessing both an  unstable and a stable photon sphere and reproducing black-hole-like external observables; and (iii) more general \qm{non-exotic compact objects}, which may or may not admit photon spheres and need not exhibit black-hole phenomenology. The log-metric can serve as an effective  \emph{geometric} candidate for the exterior spacetime of  categories (i)-(iii), and, remarkably, it is not excluded by current EHT data. These results further strengthen the physical grounding of the aforementioned scheme for NEC-satisfying metrics, whose full potential remains to be explored. Being general and  widely applicable, this method can potentially generate, beyond the specific case \eqref{eq:metric_log}, a broad family of effective geometries able to describe new prototypes of compact bodies.

In the spirit of Ref.~\cite{Li:2012zx}, the logarithmically corrected model  may also be relevant for characterizing dark matter profiles that satisfy the standard energy conditions, although the underlying effective matter content is anisotropic, while  cold dark matter is usually described as isotropic, pressureless dust~\cite{Mukhanov:2005sc}. On the other hand, we do not find any clear indication that the log-geometry provides a viable  dark energy paradigm. These points deserve a careful analysis in a separate paper.

\section*{Acknowledgments}

It is a pleasure to thank Yu-Sen An for helpful discussions. The work of Z. W. is supported by the National Natural Science Foundation of China  (12405063). E. B. acknowledges the support of INFN  {\it iniziativa specifica}  Moonlight2.

\begin{appendix}

\section{Null energy condition in non-static and spherically symmetric geometries}\label{Sec:Appendix}

Non-static and spherically symmetric metrics can be written as~\cite{Weinberg1972,Carroll2004}
\begin{align}\label{eq:metric_nonstatic}
{\rm d}s^2= - B(r,t) {\rm d}t^2+A(r,t) {\rm d}r^2 + r^2 {\rm d}\Omega^2\,.
\end{align}
 
In this geometry, the Ricci tensor $R_{\mu \nu}$ generally loses its diagonal structure, and   consequently the identification of an appropriate orthonormal frame that would diagonalize the energy-momentum tensor becomes  significantly more involved. Nevertheless, for our purposes, we can simplify the analysis by restricting attention to radial geodesics.  In this case,  the radial four-velocity takes the form 
\begin{align}
    \mathscr{V}^{\mu} = \left( \frac{\tilde{E}}{B(r,t)}, \;\mathscr{V}^{r},\;0,\;0 \right)\,,
\end{align}
with 
\begin{align}\label{Vradial_tdepend}
    (\mathscr{V}^{r})^2=\frac{1}{A(r,t)}\left(\frac{\tilde{E}^2}{B(r,t)}-\alpha \right)\,,
\end{align}
and, like before, $\alpha=1$ for timelike geodesics  and $\alpha=0$ for null ones. Here, we have  defined 
\begin{align}\label{eq:nonstaticE}
    \tilde{E}(r,t)\equiv B(r,t)\frac{\dd t}{\dd \lambda},
\end{align}
which is no longer conserved along the geodesics, as $\partial/\partial t$ is not a Killing vector. 

The energy density measured by a  radial timelike experimenter is given by  
\begin{subequations}\label{eq:WEC_nonstatic}
\begin{align}\label{eq:WEC_nonstatic-1}
T_{\mu \nu} \mathscr{V}^{\mu} \mathscr{V}^{\nu} =&\frac{1}{8 \pi } \left( \frac{{f_0}}{r^2}+\frac{{f_2} {\tilde{E}}^2}{r^2} +\frac{{f}_3\tilde{E}}{r^2} \right) \,,
\end{align}
where $f_0(r,t)$ attains the same form as  in  Eq. \eqref{eq:deff0} and 
\begin{align} \label{eq:WEC_nonstatic-3}
{f_2}(r,t):=&-r\left(\frac{1}{AB}\right)'\,,\\  
{f}_{3}(r,t):= &\frac{2r\dot{A}}{AB}\left(\frac{{\tilde{E}}^2/B-1}{A}\right)^{1/2}\,,
\end{align}
\end{subequations}
the dot denoting partial differentiation with respect to $t$. 

In analogy with the investigation in Sec.~\ref{subsec:AB vanish}, Eq.~\eqref{Vradial_tdepend} indicates that an infinite potential well may emerge where $AB$ vanishes, provided $\tilde{E}\neq 0$. Furthermore,  the energy density may diverge as the radially infalling observer approaches the location where $AB=0$. 

The NEC  for a radial null vector requires that (cf. Eq. \eqref{NEC-vectors})
\begin{align}\label{eq:nec-nonstatic}
    \frac{{f_2} {\tilde{E}}^2}{r}+\frac{2\tilde{E}\dot{A}\left(\frac{{\tilde{E}}^2}{AB}\right)^{1/2}}{AB}\geq 0\,,
\end{align}
which in the region where $AB > 0$ and $\tilde{E} > 0$ (recall that $\tilde{E} > 0$   outside the event horizon of a black hole spacetime),   boils down to  
\begin{align}
    (\sqrt{AB})'\geq \dot{A}\,.
\end{align}

For the specific case where $A(r,t) = 1/B(r,t)$, the NEC  imposes the following constraint:
\begin{align}
    \frac{{\tilde{E}}{|\tilde{E}|}\dot{B}}{B^2}\leq 0\,,
\end{align}
which in the domain where $\tilde{E} > 0$  reduces to the requirement
\begin{align}\label{eq:dynamic-NEC-Bdot}
    \frac{\partial B(r,t)}{\partial t}\leq 0\,.
\end{align}

Remarkably, a closely related monotonicity requirement also emerges in the analysis of the null convergence condition when ingoing Eddington-Finkelstein coordinates are employed, as shown  in Ref.~\cite{Borissova2025b} (see Eq.  (10) therein).

\end{appendix}

\bibliographystyle{JHEP}
\bibliography{references}{}

\end{document}